# A programmable superconductor created by light.

Viktoria Yursa[3,2], Igor Vaskivskyi[1,3], Anže Mraz[1], Damjan Svetin[1,3], Sergej Ražnjević[1], Vinko Sršan[1], Sašo Šturm[1], Tomaž Mertelj[1,3], Mikhail Feigel'man[1,3] and Dragan Mihailovic[1,2,3]

[1] *Jozef Stefan Institute, Jamova 39, 1000 Ljubljana, Slovenia*

[2] *Faculty of Mathematics and Physics, University of Ljubljana, Jadranska 19, Ljubljana, Slovenia*

[3] *CENN Nanocenter, Jamova 39, 1000 Ljubljana, Slovenia*

**The quest for superconductivity created by light extends for more than half a century[1–9], yet direct evidence of a true zero-resistance state - whose macroscopic quantum phase coherence is both created and controlled by light - has remained elusive. Here we report for the first time on a complex but robust light-programmable superconducting (LiPS) state at an aluminium-silicon heterojunction that is created and fully controlled with femtosecond laser pulses. The superconducting critical temperatures - ranging from 1.8 ~ 8.5 K, can be increased or erased at will by the application of tailored pulse sequences. At low temperatures the LiPS state shows features characteristic of a Berezinski-Kosterlitz-Thouless topological transition, but another distinct state appears at temperatures above ~2 K, which shows clear signatures of quantum phase disorder. In the presence of a magnetic field we observe behaviour characteristic of vortex pinning and creep consistent with the 2-dimensional (2D) nature of the phase coherent system. The origin of the LiPS effect is attributed to light pulse control of the Moiré-like superlattice of misfit dislocations (MDs) arising from discommensurations between the Al and Si lattices which is visible by high-resolution electron microscopy. We show how light pulses can be used to control the superlattice periodicity and highlight the appearance of topologically protected soliton-like kinks along the dislocation lines, important for imparting controllable metastability to the system. The demonstration of LiPS paves the way for designing metastable superconducting devices with controllable phase-coherence, enabling applications such as light-engineered quantum circuits, local gap tuning in quantum processors, and optically switchable superconducting devices.**

Light-induced superconductivity has long been a theoretical possibility: light can interact with the relevant excitations. Various proposals have involved pairing bosons[10–12], quasiparticles that form Cooper pairs[7], with pre-formed pairs themselves[13] or direct control via coupling with the superconducting order parameter[14,15]. Yet experimental evidence for a true long-range coherent quantum state created by light that can be universally accepted to be a superconductor has been very elusive. While many imaginative experiments have shown transient effects suggestive of superconducting behaviour—such as transient drops in electrical resistance and magnetic field expulsion in cuprates[4,6], or infrared spectral signatures in $MgB_2$,[7] and $K_3C_{60}$[8], these effects were either short-lived[4], indirect [5,8,16], or could not uniquely ascertain that a true superconducting state is present. Light-dose-dependent oxygen ion (re)ordering in underdoped high-temperature superconducting cuprates with higher oxygen content was found to lead to a metastable increase of resistive $T_c$, where photodoping[15,17,18] and/or reduced electron scattering[9,17,18] are thought to be contributing mechanisms for the enhancement of $T_c$, but this is not a fundamentally new state created by light. Recent approaches, like cavity experiments offer promising control over superconductivity[19] but have yet to deliver a clear, lasting light-induced macroscopically phase-coherent zero-resistance state[20,21]. Some more success was achieved by electric field induced superconductivity at $LaAlO_3/SrTiO_3$ interfaces[22] for example, reaching $T_c$s in the 200 mK range, and various electrostatic gating techniques in various monolayers, reaching $T_c$s in the 40 K range. But these are not switchable by light.

Here we show that a fundamentally new and persistent superconducting state with a high critical transition temperature can be created in an aluminium-silicon heterostructure and controlled with remarkable fidelity by manipulation of misfit dislocation (MD) interfacial defects with ultrashort light pulses. Once created, the light-induced superconducting state shows properties reminiscent of a Berezinski-Kosterlitz-Thouless (BKT) transition. Full coherence is ascertained with conventional 4-probe resistance, critical magnetic field ($H_{c2}$), voltage-current characteristics and critical current measurements. Remarkably, full 'Write' (W) and 'Erase' (E) control of the superconducting state is reached by individual pulses or a tailored sequence of pulses. Using combined atomic-resolution scanning/transmission electron microscopy (STEM/TEM), we observe a rectangular superlattice of MDs at the Al-Si interface - reminiscent of a Moiré pattern, arising from a discommensuration between the Al and Si crystal lattices. High-resolution microscopy shows that the MDs exert a periodic strain on both sides of the interface, introducing a quasiperiodic structural modulation enhancing the electron-phonon interaction. We show how the MD superlattice can be manipulated by light pulses, correlating MD ordering with light-induced metastable phase coherence and high $T_c$ superconductivity. We also show that in addition to MD superlattice disorder, kink defects along MD lines are very responsive to short laser pulses. These kinks – as topologically protected objects, impart metastability to the system, linking them to the experimental observation of 'programmable' superconductivity.

## Creating a superconductor with a single pulse of light

The experiments shown schematically in Fig. 1a-b were performed on nanofabricated 4-probe shaped devices made with $t$ ~15 nm Al thin films deposited through a mask onto etched Si (100) substrates. A typical device is shown in the scanning electron microscope (SEM) image in Fig. 1c. (See Methods for fabrication details and interface analysis, and supplementary information (SI) for description of electrical measurements). Prior to light exposure, the Al films exhibit typical low-temperature residual resistivity of $\rho(0) \simeq 2 \times 10^{-8}\ \Omega\ m$. The devices were exposed to 220 fs pulses from a Yb-KGW crystal laser beam focused onto a 200 $\mu m$ diameter spot covering the entire device shown in Fig. 1c). The laser wavelength of 1035 nm (1.2 eV) is close to the Al metal inter-band absorption resonance[23] (peaking at ~1.4 eV), but well below the Si inter-band absorption threshold. Bulk aluminium reflects ~95% of incident light at this wavelength. However, the optically thin Al films combined with a transparent Si substrate create an optical cavity, trapping the absorbed light within the Al layer, and effectively enhancing light intensity at the buried Al-Si interface. The calculated power density profile $P$ inside the cavity is shown in Fig. 1d. The temporal evolution of the electron and lattice temperatures, $T_e(x,t)$ and $T_L(x,t)$ respectively within the cavity calculated with a two-temperature model are shown in Fig. 1e and f) (see SI for calculation details).

The basic light-induced superconductivity phenomenon is demonstrated with an exposure of the device at 1.5 K to a *single* 220 fs pulse, which results in the creation of a superconducting state with zero-resistance at $T_c^{(1)} = 1.9{\sim}2.1\ K$. (The $T_c$ of pristine 15 nm thick Al films is ~1.19 $K$). The resistivity $\rho(T)$ as a function of temperature is shown in Fig. 1g and on an expanded scale in the insert to Fig. 1 i). We label this state SC1. Apart from a sharp drop of $\rho(T)$ to zero at $T_c^{(1)}$ we also note that the normal state resistivity $\rho_N$ drops after each laser exposure, and eventually saturates at ~ 0.6 $\rho_N$, implying that exposure to short light pulses leads to reduced electron scattering in the normal state. STEM/TEM image analysis of films exposed to femtosecond light pulses shows that the systematic drop in $\rho_N$ may be attributed to Al grain growth, consistent with reduced electron scattering within the Al film. (A detailed microscopy study is presented in the SI). A single pulse exposure at $T = 10K$ leads to a similar appearance of a resistance drop, but in this experiment true zero resistance was not observed to the lowest temperature reached by our apparatus (1.45K) (see Fig. 1h and insert to Fig. 1i).

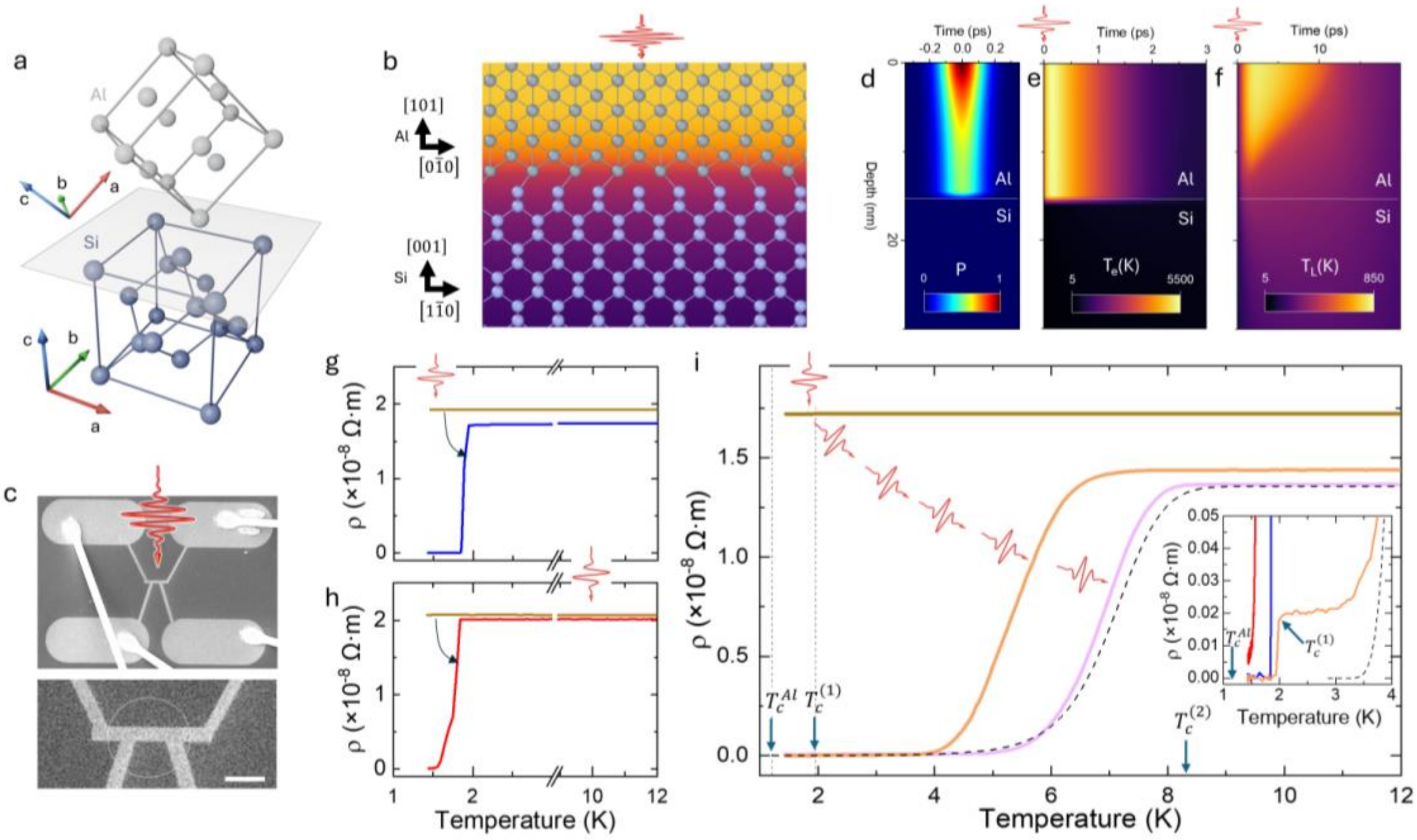

**Figure 1. Switching to a superconducting state by light pulses.** a) A schematic diagram of the Al and Si unit cell orientations at the heterostructure interface. b) an idealized cross-section of the interface indicating the lattice mismatch. The colour indicates the temperature on the same scale as f). c) The SEM images show the 4-probe device geometry with Al-wire-bonded Al contacts. The length of the bar is 20 $\mu$m. d) The absorbed energy profile within the Al film on Si vs. time. e) and f) The calculated electron and lattice temperature profiles $T_e(x,t)$ and $T_L(x,t)$ respectively as a function of time for maximum fluence. g) $\rho(T)$ measured before (brown line) and after (blue line) a *single* 200 fs light pulse at 1.5 K. $\rho(T)$ in the range 1.4 K - 300 K is shown in the SI. h) Same as g), except the single pulse is applied at 10 K. i) Resistivity $\rho(T)$ as a function of temperature before photoexcitation (brown), after 1000 pulses 50 ms apart (orange), and after a "training sequence" of 1000-pulse bursts (pink). The insert shows $\rho(T)$ on an expanded scale for the curves in g), h) and i). The line colour code is the same throughout. The dashed lines are model curves for a disordered superconductor (see text).

## Manipulating a high-$T_c$ superconducting state with light

Exposure of the device to a 'training' cycle (a burst of a thousand 220 fs pulses 50 ms apart) shifts the onset of the resistivity drop to much higher temperatures and leads to the creation of a different state labelled SC2 (Fig. 1i, orange curve). As we shall see later, SC1 and SC2 are distinct superconducting states that coexist and can be distinguished below $T_c^{(1)}$ by the application of a magnetic field. The maximum *onset* temperature increases to $T_c^{(2)} \simeq 8.5 \pm 0.2\ K$ after several training cycles (pink curve in Fig. 1i). Below ~5 K, the resistance of this state drops very close to zero. On an expanded scale shown in the insert to Fig. 1i we see that in the remanent 8 mT magnetic field of our cryostat there is a small residual resistivity of $\rho_R \sim 2 \times 10^{-10}\Omega$ m until we reach $T_c^{(1)}$. Such a residual resistivity is usually attributed to spatial gap disorder and unpaired vortices that can be moved by the passing current. It can also arise from electrical noise in the measuring circuit or the remnant magnetic field. The non-zero $\rho_R$ indicates that the phase coherence of the SC2 state is quite sensitive to external electro-magnetic perturbation. Below $T_c^{(1)}$, $\rho(T)$ of the device in the SC2 state drops to true zero as shown in Fig. 1 i). The three resistivity curves in Figs. 1 g), h) and i) are compared on an expanded scale in the insert to Fig. 1i (same colour coding). We note that the exact shape of the $\rho(T)$ curves, $T_c^{(2)}$and the value of $T_c^{(1)}$ where resistance drops to zero, and the value of the residual resistivity $\rho_R$ shown in Fig. 1 depend on the light-exposure history of the sample. The light-induced states described in Fig. 1 are stable on the laboratory timescale if the sample is kept below ~50K. Keeping a switched device at 300 K for an extended period (more than a week) causes a disappearance of superconductivity. There is no recovery of $\rho_N$ however, indicating that two independent mechanisms are operative for the drop in normal state resistance and the LiPS effect (see Si for detailed discussion of $\rho_N$). The observed S-shaped $\rho(T)$ curves in the SC2 state can be understood if we assume that the system is spatially inhomogeneous. We can reproduce $\rho(T)$ using an effective medium approximation model (EMA) of coupled superconducting puddles embedded within a conducting medium (dashed line in Fig. 1 i). The resistivity in the range $T_c^{(1)} \ll T < T_c^{(2)}$ can thus be interpreted to be a result of inhomogeneity (Fig. 3b))[24]. (A more elaborate discussion will be presented later in the paper when we discuss the detailed mechanisms, while the SI gives the detailed model calculations).

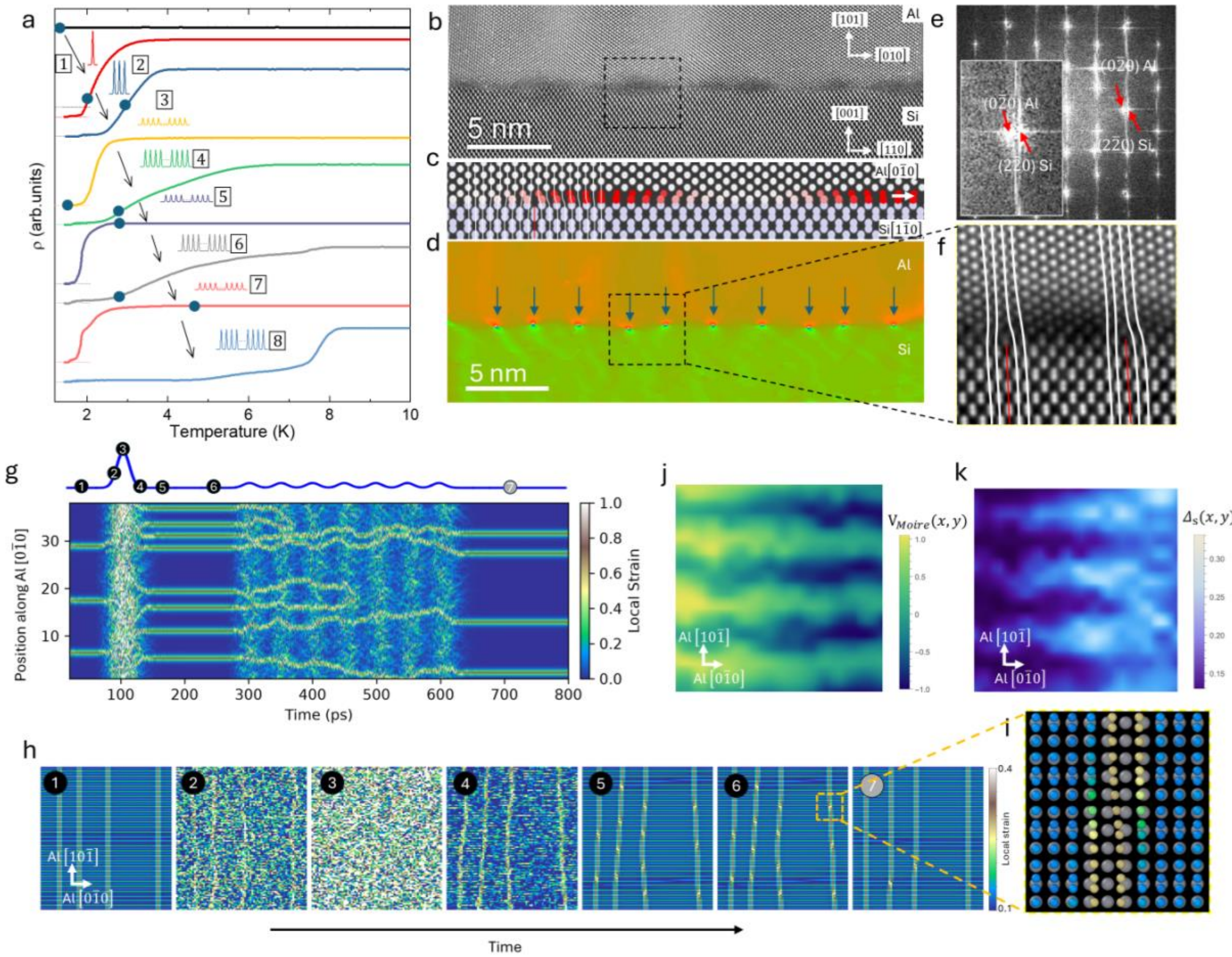

**Figure 2. Manipulating superconductivity with light pulses.** a) A progression of $\rho(T)$ curves measured after a series of light pulse sequences sequentially raise and lower $T_c$. The dot indicates the temperature at which the light is applied and the pulse sequences are as indicated. The incident fluences are (in mJ/cm$^2$): 1 : 50 (1 pulse), 2 : 50 (3 pulses), 3 : 10 (1000 pulses), 4 : 25 (1000 pulses), 5 : 10 (1000 pulses), 6 : 25 (1000 pulses), 7 : 10 (1000 pulses) and 8 : 30 (1000 pulses) respectively. Note than only ~3% of the incident light is absorbed. b) A cross-section of the Al-Si interface imaged by HAADF STEM shows a periodic structural modulation due to MDs along the $[0\bar{1}0]_{Al}$ direction (zone axis $[10\bar{1}]_{Al}$). The orthogonal direction is shown in the SI. c) An atomistic model simulation of MDs at the interface shown in b)[25]. d) The strain map from b) along the $[10\bar{1}]_{Al}$ zone axis reveals periodic MDs at the interface (arrows) with a period $\Lambda_{MD} = 2.78 \pm 0.27$ nm. Note the strain that extends well beyond the interface in the perpendicular direction. e) A Fourier transform of the HRTEM image along the $[10\bar{1}]_{Al}$ and $[110]_{Si}$ zone axis showing the different periodicities of the Si and Al lattices. f) A magnified HAADF-STEM image of two MDs indicated by dashed squares in b) and d). g) A F-K model simulation of the time-evolution of MDs along the $[0\bar{1}0]_{Al}$ direction in response to a single W/E pulse sequence (shown by the blue time-line). A proliferation of MDs is observed after a single short laser pulse, creating the metastable LiPS state. The pattern is erased by a subsequent train of weaker pulses, causing the system to relax to a different discommensuration pattern. h) A 2D 'top view' simulation of the MD strains at the interface responding to a 10 ps W pulse with F-K model parameters similar as in g). The images are numbered as shown in the time-line in g). Kinks appear as topological defects at MD junctions (yellow cross-wise stripes). i) The kink structure on an expanded scale. j) The Moiré potential ($V_{Moire}(x, y)$ map and k) the corresponding superconducting gap map $\Delta_s(x, y)$ for a system with MD discommensuration disorder potential shown in h). Note that the large gap (pairing density) regions in $\Delta_s(x, y)$ correspond to valleys in $V_{Moire}(x, y)$. (See Si for details of model calculation details presented in this Figure).

Remarkably, both $T_c^{(1)}$ and $T_c^{(2)}$ can be controllably increased *or decreased* by applying appropriate write (W) and erase (E) pulse sequences. In the example shown in Fig. 2a, first a single W pulse creates a state with $T_c \simeq 2$ K. After three such pulses $T_c^{(2)}$ increases, but after exposure to a weaker 1000 light-pulse train, $T_c^{(2)}$ is reduced back to < 1.7 K. Repeating W/E cycles, we reach $T_c^{(2)} \simeq 8.1$ K. Conversely, the application of multiple E pulse trains leads to complete erasure of all traces of superconductivity to <1.45 K. The shape of the $\rho(T)$ curves varies from cycle to cycle and suggests that the superconducting state is inhomogeneous and the W/E procees is not entirely deterministic, which - as we shall discuss later - gives important clue to the origin of the effect. Such behaviour is qualitatively completely reproducible and has been observed in multiple devices, but the shape of the $\rho(T)$ curves vary across different cycles and devices (see SI for various examples and stability demonstrations).

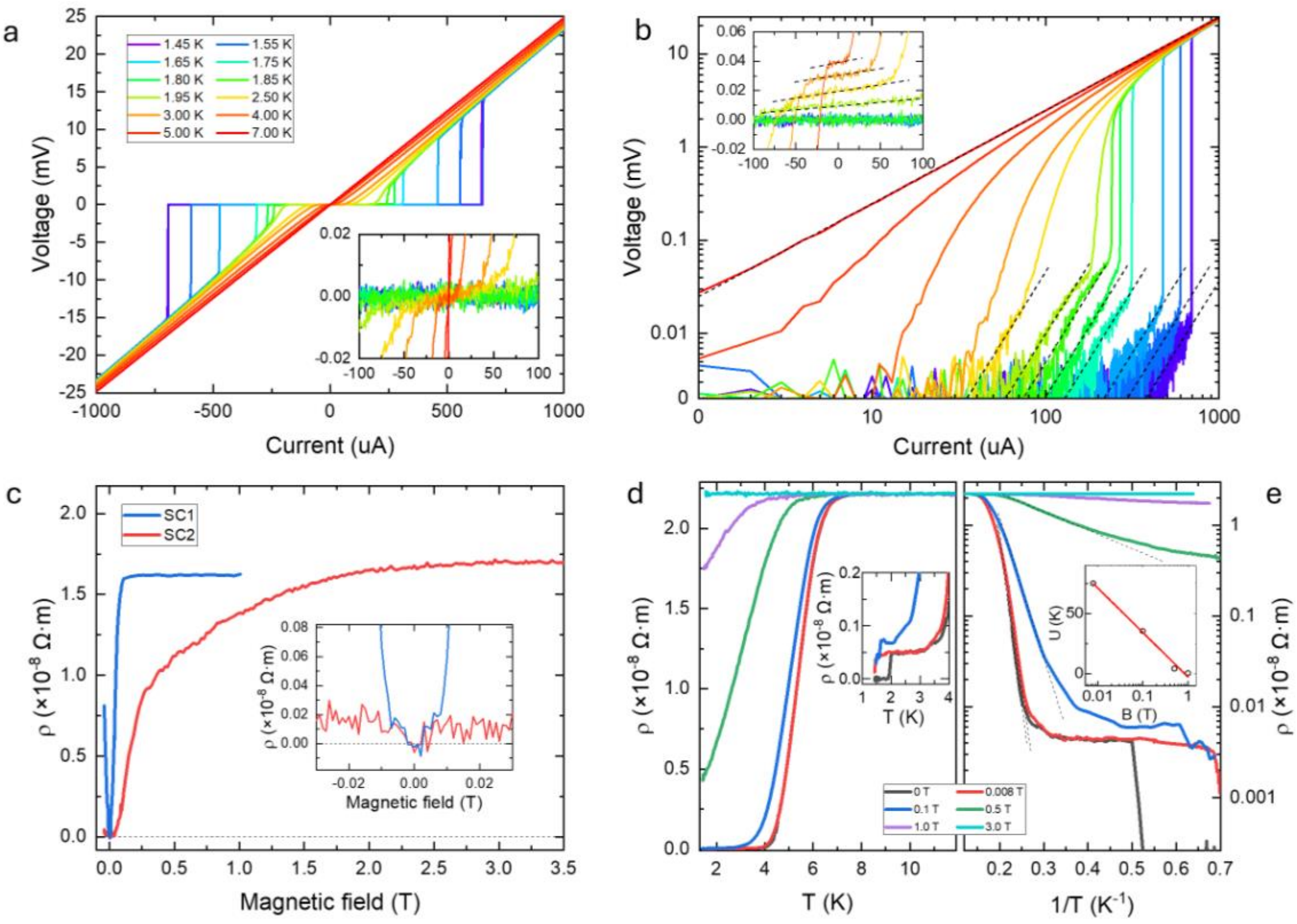

**Figure 3. Properties of the light-induced superconducting state.** a) V-I characteristics of the SC2 state at different temperatures. The insert shows the linear V-I behaviour at the origin. b) V-I curves of the SC2 state on a log-log plot show power law behaviour $V \sim I^{\alpha}$, with $\alpha = 3$ for $T \rightarrow T_c^{(1)} \simeq 2.0\ K$ from below. The dashed lines indicate a slope with $\alpha = 3$ *below* 2K. The insert shows offset curves *above* 2 K on a linear scale showing $\alpha = 1$ for $T > T_c^{(1)}$. c) Resistivity $\rho(B)$ as a function of magnetic field at $T = 1.5K\ (< T_c^{(1)})$ in the SC1 state (blue) and the SC2 state (red) respectively. The upper critical fields are $H_c^{(1)} \simeq 120\ mT$ and $H_c^{(2)} \simeq 3\ T$ respectively. The insert shows the true zero resistance near $B = 0$. d) $\rho(T)$ measured in different magnetic fields (left). The insert shows the behaviour near the origin. e) An Arrhenius plot of $\rho(T)$ vs. 1/T shows a cross-over from thermally activated to T-independent resistivity consistent with a crossover to MQTV at low T. The dotted lines are Arrhenius fits. The insert shows the MQTV barrier energy U as a function of B.

## Characteristics of the light-induced superconducting state

To understand the origin of the light-induced state, we first analyse its superconducting properties. The $V - I$ curves measured over a range of temperatures on a LiPS state with $T_c^{(1)} = 1.9K$ and $T_c^{(2)} \simeq 5\ K$ (Fig. 3a) display some features commonly observed in 2D superconductors. At 1.45 K the critical current is $I_c \simeq 700\mu A$ (Fig. 3a)), which is comparable to some known interfacial superconducting systems[24]. A recognizable feature of a 2D topological transition characteristic of BKT superconductors is power law behaviour $V \propto I^3$ as $T \rightarrow T_{BKT}$ from below, where $T_{BKT}$ is the BKT transition temperature which arises from vortex-antivortex unbinding by the Lorentz force caused by the passing current[26].

Examining a log-log $V - I$ plot in Fig. 3 b) we see that as $T \rightarrow T_c^{(1)}$, in the temperature range $1.45K < T \leq 1.95$ we observe distinct power-law behaviour of the form $V \sim I^{\alpha}$ where $\alpha \simeq 3$ over a range of currents. Unusually, however, the power law behaviour does not extend all the way to $I_c$. Instead, we observe an abrupt voltage jump close to $I_c$, which is more characteristic of a bulk superconductor than a BKT transition close to $I_c$. Immediately above $T_c^{(1)}$ at *low currents*, the V-I curve is linear (shown in the insert to Fig. 3b), with $\alpha = 1$ from 1.95 to ~3 K, consistent with a BKT transition[24] at $T_c^{(1)}$. For larger currents $\alpha$ increases rapidly, but it is difficult to identify any regions in the I-V curves that show constant $\alpha$ for temperatures up to $T_c^{(2)}$. The system thus shows complex multi-component behaviour with an anomalous BKT-like transition associated with $T_c^{(1)}$, and rounded V-I curves associated with $T_c^{(2)}$ consistent with an inhomogeneous system.

**Magnetic field measurements** reveal some further unusual properties of the light-induced superconductor, and address the existence of a true superconducting state, which is characterised by $\frac{dV}{dI}|_{I=0} = 0$ for $H > 0$ as $T \rightarrow 0$. Fig. 3 c) shows $\rho(B)$ as a function of applied magnetic field $B$ in the SC1 and SC2 states measured separately at 1.5 K. The SC1 state ($T_c^{(1)} = 1.9K$) displays an upper critical field of $H_{c2}^{(1)} \sim 120\ mT$. In contrast, after further light pulse exposure the same device into the SC2 state (red curve in Fig. 3c)) measured at 1.5 K shows a slow saturation and a significantly higher field at saturation ($H_{c2}^{(2)} \sim 3\ T$). This clearly shows that the SC1 and SC2 states can be distinguished below $T_c^{(1)}$ by their distinctly different behaviour in magnetic field. Assuming a homogenous type II superconductor for SC1, the vortex size in the SC1 state can be estimated from the Ginzburg-Landau formula for the coherence length $\xi^{(1)} = \sqrt{\phi_0/2\pi H_{c2}^{(1)}} \simeq 50\ nm$, where $\phi_0 = h/2e$ is the flux quantum. We note that $\xi^{(1)}$ is still much shorter than the coherence length of bulk Al ($\sim 1\ \mu$m), but longer than the Al film thickness $t = 15$ nm. In the SC2 state the gradual saturation- which we attribute to inhomogeneity, prevents us from accurately determining the upper critical field. However, the observed saturation field of ~3 T implies a much shorter coherence length of $\xi^{(2)}$<10 nm.

A plot of $\log(\rho)$ vs. 1/T (Fig. 3e)) in the SC2 state shows a crossover from temperature-dependent Arrhenius-like behaviour of $\rho$ at high temperatures to T-independent resistivity at low temperatures where the activation energy is field-dependent. Such behaviour has been previously attributed to macroscopic quantum tunnelling of dislocations in the vortex lattice (MQTV).[27] Alternatively, it could be attributed to external circuit noise, which would result in similar T-independent behaviour[28]. The dependence of the activation energy $U$ shown in the insert to Fig 3e) is consistent with the relation[29] $U = U_0 ln(H_{c2}^{(1)}/H)$, which characterises the collective creep regime reported previously in highly disordered thin films[27], where $U_0 = \phi_0^2 t/256\pi^3 \mathrm{L}^2$, $t$ is the film thickness and L is the penetration depth. A fit to the $U$ vs. $1/T$ plot (shown in the insert to Fig. 3e), gives $U_0 = 1.38 \pm 0.1\ meV,$ and $H_{c2} = 1 \pm 0.2$ T, consistent with Fig. 3c. From $U_0$ using the relation[30] $\Theta(0) = \frac{\phi_0^2 d}{16\pi^2 \Lambda^2} = 16\pi U_0$ we can roughly estimate the energy scale associated with the 2D zero-temperature superfluid stiffness to be $\Theta(0) \approx 69 \pm 5\ meV$. Another estimate of the the superfluid stiffness $\Theta$ per square for the SC1 state can be obtained from the critical current in Fig. 3a using the relation[31] $j_c = \frac{4e}{3\sqrt{3}\hbar} \frac{\Theta(T)}{\xi}$ where $e$ is the electron charge and $\xi$ is the superconducting coherence length, the critical current density per unit length of the film $j_c = I_c / l_\perp$ (where $l_\perp$ = 10 µm is the film width). With $j_c$ = 0.7 A/cm at the (sufficiently low) temperature T = 1.45K and $\xi^{(1)} \simeq$ 50 nm, one finds $\Theta \approx 20$ meV. Both estimates of $\Theta$, from $U_0$ (69 meV) and from $J_c$ (20 meV) are much larger than the BCS superconducting gap for either state: $\Delta_{BCS}^{(1)} = 1.75 k_B T_c \approx$ 0.3 meV, and $\Delta_{BCS}^{(2)} \approx$ 1.3 meV respectively. Such discrepancies are usually observed in superconductors with a moderate amount of disorder[32], and are consistent with the S-shaped resistance curve below $T_c^{(2)}$.

## The origins of light-programmable superconductivity (LiPS).

The most non-trivial property of the light-induced state is the fully controllable W/E cycling (Fig. 2), which is also the key to understanding the mechanism. Firstly, the observed reversibility behaviour rules out processes which are not time-reversible, particularly bulk interdiffusion and alloy formation. This is relevant because bulk $Al_{1-x}Si_x$ non-equilibrium alloys show superconducting behaviour with $T_c$s peaking at ~8 K for $x = 0.3{\sim}0.6$ [33–38] that could possibly point toward the origin of the observed high-$T_c$ superconductivity. Such alloy formation by light is limited to the interface layers by the short timescale of the laser pulse excitation and the high activation energies for diffusion (0.8~1.2 eV for Si in Al[39,40], and 3.4 eV for Al in Si respectively[39,40]). Calculations show that for a single laser pulse the Si diffusion into Al is limited to < 0.1 nm and is substantially smaller for Al in Si (see SI). The absence of alloying and interdiffusion is confirmed by EDS and EELS experiments (see SI). Additionally, $Al_{1-x}Si_x$ alloy formation causes a rapid increase of normal state resistance with $x$[33–38], but we observe the opposite (Fig.1). Nevertheless, the fact that we observe grain growth by HRTEM indicates short range diffusion does take place at grain boundaries in the Al film, which is most likely driven by the strains which are clearly visible in Fig. 2d. Other known mechanisms for raising $T_c$ in Al, such as granularity[42], confinement[43] or reduction of electron scattering[18] are not compatible with the observed W/E systematics. Finally, enhanced electron-phonon coupling under high pressure in metallic hexagonal Si was considered as responsible for a superconducting transition at 8.2 K[41], but such a structure of Si is very different to what we observed and shows no evidence of metastability, so this can also be excluded.

Having considered some possible conventional mechanisms, we focus on the most important effect at the interface, namely the appearance of periodic misfit dislocations (MDs) arising from the difference in lattice periodicities of Si and Al surfaces. The HAADF-STEM and HRTEM studies of the interface cross-section shown of the device in Figs. 2 b, d and e, f reveal periodically spaced MDs located regularly at the interface with a period of $\Lambda_{MD}^{[0\bar{1}0]}$ =3.6 $\pm$ 0.5 nm along the Al [0$\bar{1}$0] direction. In the perpendicular direction the dislocation density is higher due to larger misfit of ~25% (see SI for HAADF images). The substantially different thermal expansion coefficients of Al and Si lead to temperature-dependent rearrangements of MDs, and it is known experimentally that MD motion in Si interfaces is thermally activated[44]. The MDs thus form a temperature-tuneable 2-dimensional rectangular superlattice of discommensurations that can be controlled by the thermal perturbation caused by the laser pulses. The dynamics of this process can be visualized by considering the time-evolution of discommensurations in response to the thermal perturbation. The total energy state along each spatial axis can be modelled by the classical Frenkel-Kontorova (F-K) model Hamiltonian[45,46] which includes the harmonic spring interactions of the epilayer and the periodic potential field modulated by the Si substrate:

$$H = \sum_i \left[\frac{1}{2} m(\dot{u}_i)^2 + \frac{1}{2} k(u_{i+1} - u_i - a_{\mathrm{Al}})^2 + V_0\left(1 - \cos\left(\frac{2\pi u_i}{a_{\mathrm{Si}}}\right)\right)\right] \quad (1)$$

Here $m$ is the atomic mass, $k$ represents the intra-layer spring elastic modulus, $V_0$ is the Peierls-Nabarro substrate potential depth, $a_{\mathrm{Si}}$ is the background substrate period, and $a_{\mathrm{Al}}^{\alpha} = a_{\mathrm{Si}}(1 + \delta_\alpha)$ specifies the unconstrained natural atomic spacing of the anisotropic Al epilayer. Fig. 2 g shows a model calculation of the time-evolution of the effective strain due to discommensuration MDs along the Al$[0\bar{1}0]$ crystal axis following W and E laser pulse sequences. (The calculation details are given in the SI.) The initial regular array of equally spaced MDs is strongly distorted by the single quench pulse, resulting in an initially denser metastable configuration of MDs. This irregular pattern persists until the annealing E pulse train arrives, causing a relaxation of the strain, and eventual recovery of the original MDs density, but not necessarily with the same Moiré pattern. The recovery of the MD pattern after relaxation to room temperature is confirmed by HRTEM (see SI) and is consistent with the observed relaxation of the LiPS state at 300K.

Fig. 2 h illustrates the predicted MD dynamics after the W pulse (Fig. 2 h, panels 1-6) and a pulse train of E pulses (Fig. 2 h, panel 7) within a 2D F-K model[47], revealing the emerging disorder of the superlattice pattern after the laser pulse: the pattern is initially denser and more irregular before eventually relaxing as a result of thermal fluctuations to a different pattern than the initial one. The 2D calculation (see SI) also reveals kinks in the metastable MD lines particularly at MD crossings, which appear and disappear in pairs. These topologically protected soliton-like structures are known to impart metastability to the system[47] and prevent relaxation of the Moiré superlattice. The time-dynamics of kink density in response to W and E pulse sequences is presented in the SI. An important observation from our electron microscopy is the appearance of lattice strain which originates from the MD lines at the interface, and spreads away from it in a perpendicular direction. It is clearly visible on both sides of the interface (orange and green streaks on the Al and Si side of the interface as shown in Figs. 2 b d and f, as well as the SI for other directions along the interface).

Relating the MD dynamics to superconductivity, changing the distance between MDs directly manipulates $T_c$ through three distinct physical mechanisms: microscopic electronic confinement, macroscopic strain, and spatial phase fluctuation. A periodic structure of MDs leads to a tendency towards Brillouin zone folding and the associated creation of narrow bands[48] at the interface. Note that the strains emerging from MDs at the interface extend quite far beyond the actual boundary (red and green streaks in Al and Si respectively as shown in Fig. 2d). This reduces the electronic kinetic energy and effectively increases the density of electronic states at the Fermi energy $N(E_F)$. A model for superconductivity emerging within a two-dimensional rectangular Moiré superlattice may be constructed by constructing an effective tight-binding Hamiltonian:

$$H_0 = -t_{\mathrm{eff}} \sum_{\langle i,j \rangle,\sigma} \left( c_{i\sigma}^{\dagger} c_{j\sigma} + \mathrm{H.c.} \right) - t_{\mathrm{eff}}' \sum_{\langle\langle i,j \rangle\rangle,\sigma} \left( c_{i\sigma}^{\dagger} c_{j\sigma} + \mathrm{H.c.} \right) - \mu \sum_{i,\sigma} n_{i\sigma} \quad (2)$$

where $t_{\mathrm{eff}}$ and $t_{\mathrm{eff}}'$ represent the effective nearest-neighbor and next-nearest-neighbor the effective nearest-neighbour hopping parameters across the Moiré superlattice sites, and $\mu$ is the chemical potential. Because the MD periodicity $\Lambda_{MD}$ is much larger than either lattice constant, the Brillouin zone folding gives rise to flat minibands which effectively enhance $N(E_F)$. With phonon-driven pairing, the superconducting order parameter is local, and the critical temperature is given by: $T_c \propto \langle\omega\rangle \exp\left(-\frac{1.04(1+\lambda)}{\lambda-\mu^*(1+0.62\lambda)}\right)$. The electron-phonon interaction $\lambda$ is typically expressed as $\lambda \sim 2N(E_F)\int_0^{\infty}\frac{\alpha^2(\omega)F(\omega)}{\omega}d\omega$, where $F(\omega)$ is the phonon density of states and $\alpha^2(\omega)$ is the electron-phonon coupling matrix element. We note that if the Fermi level can be tuned - for example by carrier photodoping, close to a van Hove singularity of the Moiré lattice, this will have an effect on $\lambda$. Indeed, bond reconstruction[49] and trapping of photoexited charge carriers is quite likely[50] at the Al-Si interface, where $3s$ and $3p$ aluminium atom orbitals hybridize with the reconstructed 2×1 surface of Si(100)[51].

In the presence of disorder of the MD superlattice seen in HRTEM (Fig. 2d), and predicted by the F-K model Fig. 2g and h), we expect a spatial variation of the superconducting order in the vicinity of the interface. A model calculation illustrating the spatial dependence of the superconducting gap $\Delta_s(x,y)$ in the presence of disorder of the Moiré lattice is shown in Fig. 2j (see SI for details). Note that streaks of $\Delta_s(x,y)$ (bright regions in Fig. 2 k) align with the potential valleys/wells of $V_{\text{Moiré}}(x,y)$ (dark regions in Fig. 2 j), where local electron density accumulates to form flat-band states which contribute to pairing. Depending on the degree of disorder and the inherent anisotropy of the MD lattice, the phase coherence can be anisotropically suppressed, leading to meandering 'rivers' of superconducting order. This can be directly validated by mapping the local superconducting gap $\Delta_s(x,y)$ with scanning tunnelling microscopy for example. Furthermore, the residual resistance arising from phase slips between superconducting 'rivers' may be improved by devising optimised LiPS protocols for controlling MD disorder and the associated kink/antikink spatial distribution.

The demonstration of direct control of the MD lattice by laser pulses suggests that the metastable superconducting state is directly controllable not only by light, but may be manipulated by any external perturbation that can affect MD discommensuration order, such as charge injection or pulsed current in the normal state. The 'training' behaviour shown here is strongly reminiscent of the metastable behaviour associated with the charge density wave domain manipulation in the metastable state of 1T-$TaS_2$[52,53] that is controllable both by light and by charge injection. In both cases, the metastability is emergent from topologically arrested dynamics of topological defects - kinks at MD crossings in Al-Si interfaces, and domain wall junctions in 1T-$TaS_2$ respectively. We speculate that other superconducting thin film systems such as $La_{1.9}Sr_{0.1}CuO_4$ and $YBa_2Cu_3O_{7-d}$, where $T_c$ [54] or $J_c$ [55] are controlled by misfit strain may be candidates for the LiPS effect.

We conclude by pointing out that amongst many possible applications, LiPS introduces the possibility of dynamically controlling Al-on-Si circuits that currently form the backbone of superconducting quantum computing technology. For example, quasiparticle traps and barriers could be patterned by laser to control quasiparticle motion in superconducting circuits[40]. Mask-less direct laser-writing[56] of buried interface properties also opens possibilities for fabricating new types of quantum devices, both active and passive. The demonstration of a programmable phase coherence in a metastable macroscopic quantum state suggests that we need to revise our notion of superconductivity being limited to the ground state of quantum matter.

**Acknowledgments:**

The authors acknowledge Gabriel Aeppli, Simon Gerber, Serguei Brazovskii, Wulf Wulfheckel, Alexei Lunkin, Benjamin Sacepe and Viktor Kabanov for useful discussions and Bojan Ambrožič, Gregor Kapun, Nik Gračanin for TEM lamela preparation; and Sandra Drev for TEM characterization of the lamellas. The authors wish to acknowledge funding from ERC AdG HIMMS and ARIS funding P-0040. VY acknowledges funding via public call for scholarships for doctoral studies for foreign citizens in the field of Quantum Technology by Public Scholarship, Development, Disability and Maintenance Fund of Republic Slovenia.

**Methods:**

**Circuit preparation.** Devices designed for 4-contact measurements were made by evaporating thin Al films (typically 15 nm) through a SiC shadow mask onto (100) oriented, BOE etched Si substrates (Fig. 1a). Contacts were Al-wire-bonded directly to the Al circuit. The resistance of the device was measured with nanovoltmeter in delta-mode (Keithley 2182 A) or with a lock-in detector (Zurich instruments MFIA). Plots of resistivity $\rho$ in the range 1.4 K $< T <$ 300 K for typical circuits used in the measurements are shown in the SI. The magnetic field measurements were performed in an Oxford Instruments Spectromag superconducting magnet with a variable temperature insert, with magnetic field perpendicular to the Al film. Unless stated otherwise, all the experiments were performed in a remnant field of $B = 8$ mT.

**High-resolution transmission electron microscopy (HRTEM)** and high-angle annular dark-field (HAADF) imaging in scanning transmission electron microscopy (STEM) mode were performed using a probe-corrected Thermo Fisher Scientific Spectra 300 microscope operated at an accelerating voltage of 300 kV. HRTEM images were recorded using a CETA camera at a magnification of 570000, corresponding to a pixel size of 13.33 pm. The collection semi-angles of the HAADF detector were set to 51–200 mrad, and the dwell time was 10 µs. For strain analysis, geometrical phase analysis (Du, H. (2018) GPA – geometrical phase analysis software) was performed on the HRTEM images. Misfit dislocations along the $[10\bar{1}]_{Al}$ zone axis were identified in the HAADF images using the in-plane phase component obtained from the GPA.

Orientation mapping was performed in microprobe STEM mode using a convergence semi-angle of 1.57 mrad. A 4D-STEM dataset was acquired using an Electron Microscope Pixel Array Detector (EMPAD), with the camera length set to 460 mm. The field of view was 1.2 µm, and the scan was performed over a 256 × 64 pixel grid, corresponding to a pixel size of 4.7 nm. Post-processing of the orientation mapping data was carried out using py4DSTEM[57].

**Author contributions:** VY, IV and DM performed the majority of the work. VY and IV performed the measurements; VY and DS prepared devices; IV, VY, MF and DM performed the superconducting state analysis. TM contributed to the discussion. SS, SR and VS performed the electron microscopy analysis. Model calculations were performed by IV and DM. DM proposed the experiments and wrote the manuscript with contributions from IV, VY and MF.

# Supplementary information: methods, electron microscopy results and model calculation details

Accompanying the manuscript: *"A programmable superconductor created by light"* by Yursa et al.

Table of Contents

# 1. Electrical measurements of the Al-Si interface/Al sandwich.

The 4-probe circuit used for resistivity measurements is shown in Figure SI 1a. The contacts are wire bonded directly to the Al film using standard Al wire-bonding wire (Figure SI 1b). The geometry of the devices is designed such that either part or the entire device can be illuminated by the laser (Figure SI 1c). The diameter of the laser beam waist is typically 50~200$\mu$m.

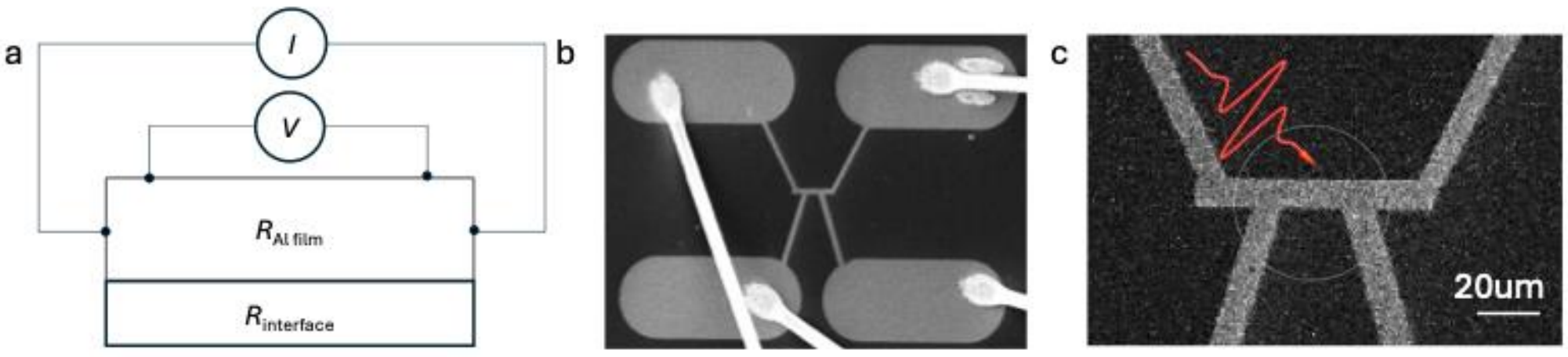

*Figure SI 1. a) Schematic diagram of the 4-probe measurement circuit. b) The contacts are bonded to the Al film. c) A 50$\mu m$ laser spot is shown schematically. The width of the strip is 10um.*

All data are taken in the dark to avoid the measurement of any photocurrent through the Si substrate. The non-illuminated undoped Si substrate is non-conducting. The 15 nm Al film is assumed to have a superconducting critical temperature close to bulk, which is $T_c$=1.19 K, well below the 1.45 K temperature limit of our apparatus. The pristine devices typically show no downturn in resistivity to 1.45 K.

We note that devices made with Al films deposited on sapphire did not show the effects we describe here for the Al-Si devices, implying that the Si substrate plays a role in the LiPS effect.

The measured system effectively consists of a 2D interface in parallel with the Al film. The (dark) normal state resistance of the Al film at low temperature (<10 K) is typically $R_{Al} = \frac{\rho_N}{t} \simeq 1\ \Omega$. Here $t$ is the film thickness and $\rho_N$ is the resistivity (see Figure SI 15 for the resistivity curves of the subset of devices for which resistivity was measured over the full temperature range from 1.4 to 300 K). This is much lower than the quantum sheet resistance of $R_Q = \frac{h}{e^2} = 25.8\ k\Omega$ and is comparable with MBE-grown films of Al on Si (100) surfaces.[1]

Any dark conductance of the interface itself is not measurable independently in the normal state because of the high conductance of the Al film. If the interface resistance drops below the Al thin film resistance (~1 Ω), it becomes visible, particularly in the case of a superconducting transition.

The hallmark signature of a superconducting metal-to-insulator (S-I) transition observed in cuprates and other 2D systems is the evolution of the resistivity from a superconductor to insulator through a critical point marked by a constant quantum sheet resistance $R_Q$ (Figure SI 2). Such a quantum critical transition might occur as a result of light exposure at the Al-Si interface, by analogy to system tuning by doping, gating[2,3], application of magnetic field[2] or variation of film thickness[4]. However, in our case the 2D properties of the Al-Si interface itself are not directly measurable because of the low resistance of the high conductance Al layer which is effectively in parallel with the interface. The resistance of the interface 2D layer is thus detectable only once it drops below $R_{Al}$ as shown schematically in Figure SI 2.

While the fast drop in resistance at $T_c^{(1)}$ in the SC1 sate is consistent with this plot, the SC2 resistance drop is more gradual, and does not follow this scenario, as it asymptotically approaches $R_N$ (See for example Fig. 1i and Fig. 2a of main text). Our data thus favour an interpretation whereby the LiPS SC2 state is a result of modification of Al film by the interfacial Moiré MD pattern.

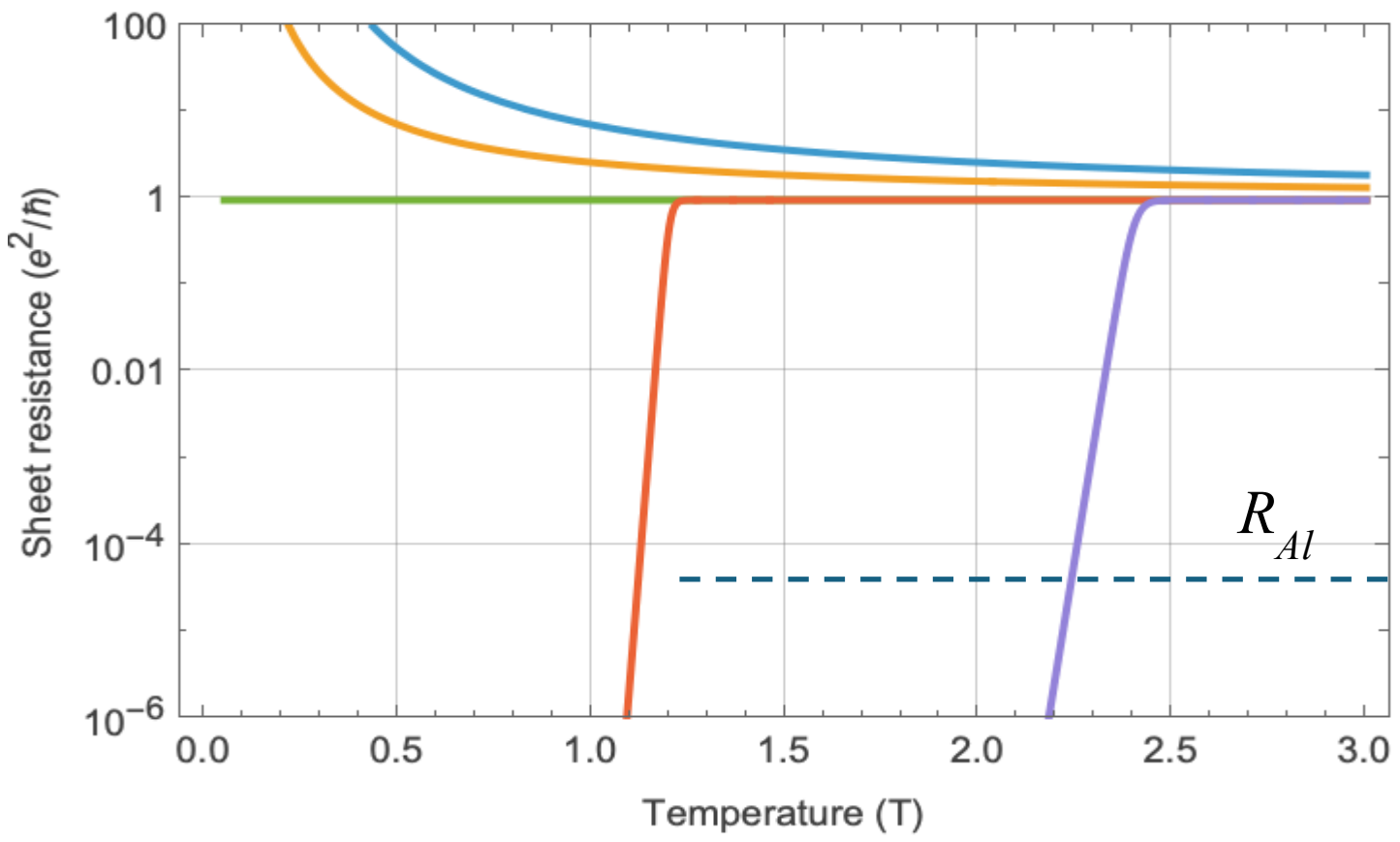

*Figure SI 2. Schematic behavior of a hypothetical sheet resistance as a function of temperature associated with a superconductor-to-insulator transition at the Al-Si interface. The dashed line shows the normal state $R_{Al}$. Al becomes superconducting below ~1.19 K (not shown). This conventional scenario is not consistent with our resistivity curves, particularly for the SC2 state.*

## 2. Thermal Dynamics:

### Two-Temperature Model (TTM)

The TTM is used to calculate the time-evolution of the electronic and lattice temperatures in response to an absorbed laser pulse. We use T-dependent thermal constants to ensure accuracy. The temperatures of the electrons ($T_e$) and the lattice ($T_l$) are governed by the following coupled differential equations:

$$C_e(T_e)\frac{\partial T_e}{\partial t} = \nabla \cdot (\kappa_e \nabla T_e) - G(T_e - T_l) + S(z,t)$$

$$C_l(T_l)\frac{\partial T_l}{\partial t} = \nabla \cdot (\kappa_l \nabla T_l) + G(T_e - T_l)$$

Where:

- $C_e(T_e)$ is the electron heat capacity.
- $C_l(T_l)$ is the lattice heat capacity.
- $\kappa_e$ is the temperature-dependent electron heat conductivity.
- $\kappa_l$ is the temperature-dependent lattice heat conductivity.
- $G$ is the electron-phonon coupling constant.

The laser source term $S(z,t)$ for a Gaussian pulse centered at $t_c$ is: $S(z,t) = \frac{P(z)}{\tau_p\sqrt{2\pi}}\exp\left(-\frac{(t-t_c)^2}{2\tau_p^2}\right)$, where $P(z)$ is the absorbed power density as calculated by transfer matrix method for two-layer system (see below).

The coupled differential equations were solved numerically for two-layer system (15-nm thick Al and 500-µm thick Si). We used calculated temperature-dependent electron-phonon coupling, electronic specific heat and heat conductivity from XTANT database[5] and tabulated temperature-dependent total specific heat and heat conductivity[6], from which the phonon parameters were estimated.

### Optical absorption profile and cavity effect in thin Al films

Since the Al film thickness (~15 nm) is well below the optical penetration depth at 1035 nm, absorption is governed by cavity effects rather than bulk attenuation. We model the depth-resolved power density using the transfer matrix method, accounting for coherent effects in Al film while treating the 500 μm Si substrate as an incoherent medium (Figure SI 3). Film thickness modulates both the total absorbed energy and the absorption depth profile. While thinner films behave as lossy cavities, even for intermediate thicknesses with low Q-factors as in our 15-nm films, back-reflection significantly enhances absorption near the Al-Si interface. The results are given in Figure SI 3.

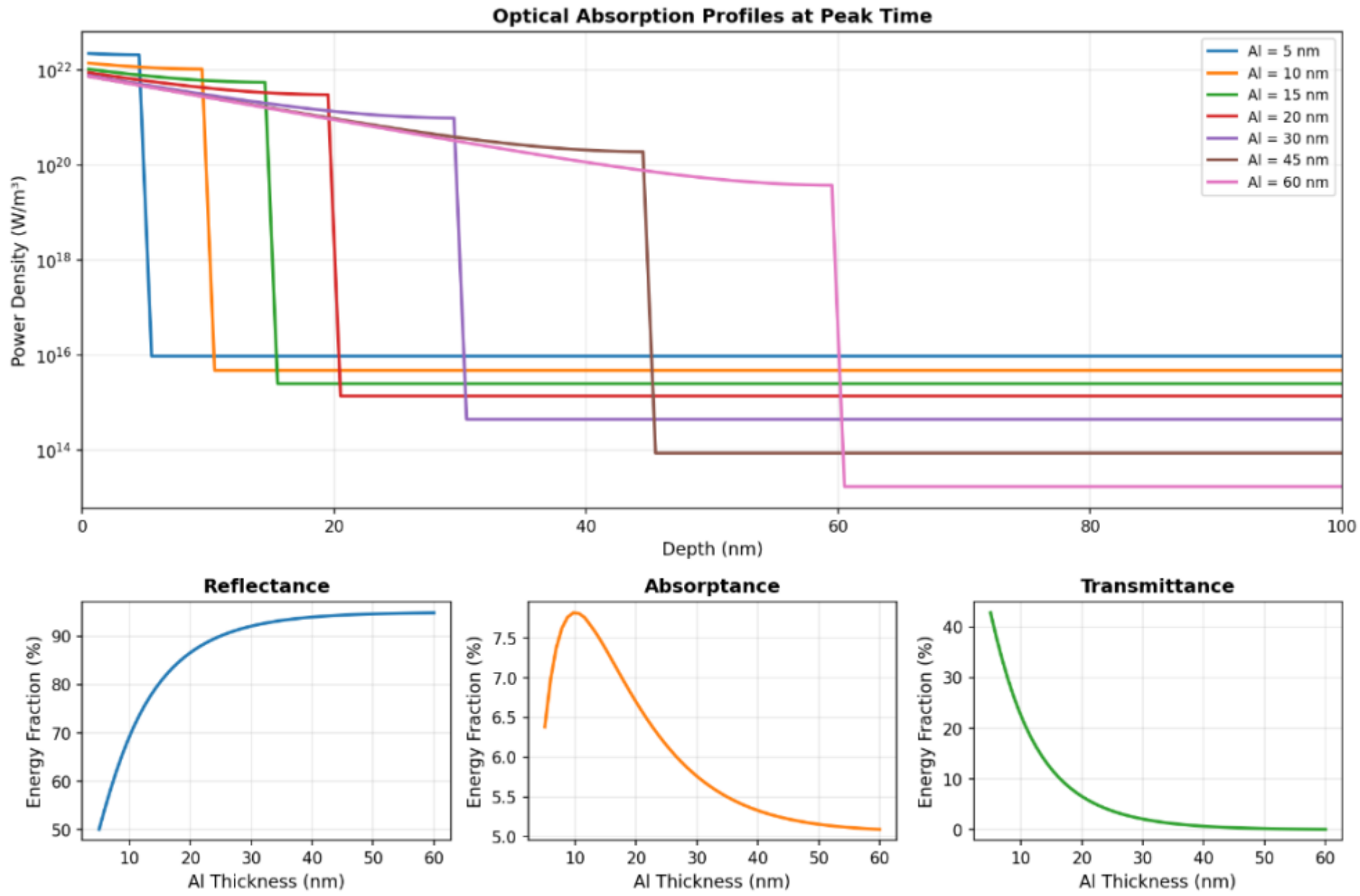

*Figure SI 3. **Cavity-enhanced optical absorption in Al-on-Si devices.** (a) Depth-resolved optical power deposition profiles for varying Al film thicknesses. (b-d) Normalized reflectance, absorptance, and transmittance of the device as a function of Al film thickness.*

### Thermal Misfit Strain calculation

The mismatch in the Coefficients of Thermal Expansion (CTE) between the Aluminum film ($\alpha_{Al}$) and the Silicon substrate $\alpha_{Si}$ induces a transient thermal misfit strain $\epsilon(t)$. Assuming a linear expansion model, the strain is expressed as:

$$\epsilon_{misfit}(t) = (\alpha_{Al} - \alpha_{Si})\,[\,T_l(t) - T_0\,]$$

Given $\alpha_{Al} \approx 23.1 \times 10^{-6}, K^{-1}$ and $\alpha_{Si} \approx 2.6 \times 10^{-6}\,K^{-1}$, the peak strain reached at $T_l \approx 950$ K is approximately 1.33%.

## 3. STEM analysis of the Al film and the Al-Si interface

### Lamella preparation

High-resolution (scanning) transmission electron microscopy (HR-(S)TEM) was used to analyze the discommensuration defects at the Al/Si interface. The lamellae for (S)TEM studies were prepared using Ga-based FIB milling. For each investigation, in addition to the lamella prepared from the active (laser-exposed) part of the device, a reference lamella was cut from the unexposed part (contact pad area) of the same device. This ensures that the exposed lamella and the unexposed lamella shared the same thermal cycling history. The analysis was made for devices that showed characteristic LiPS behavior. All FIB and (S)TEM processing was performed at room temperature.

### Low magnification STEM

In both the exposed and unexposed regions of the device, we observe well-defined granular morphology. Photoexcitation systematically lead to an increase in the average grain size, which is already visible in the raw low-magnification images (Figure SI 4 a,b). A quantitative statistical analysis confirms this trend: the mean grain diameter grows from 30 nm to 44 nm (Figure SI 4 c). We noted in the main text that the increase in grain size is consistent with our observation of systematic lowering of the normal-state-resistance after exposure to light. However, we note that the grain size systematics is inconsistent with the reversible W/E LiPS behavior of Fig. 1i and Fig. 2a of the main text.

A detailed structural analysis shows that the Al film is predominantly growing in the (101) direction, with only a low number of small grains exhibiting alternative orientations. For reference, the directions of the grain alignment are shown in Figure SI 8. Laser exposure drives coarsening and reorientation of these minority grains, resulting in a film that becomes nearly fully (101)-oriented with approximately equal number of grains oriented (0-10)[10-1]$_{Al}$ ‖ (1-10)[110]$_{Si}$ and (10-1)[010]$_{Al}$ ‖ (1-10)[110]$_{Si}$.

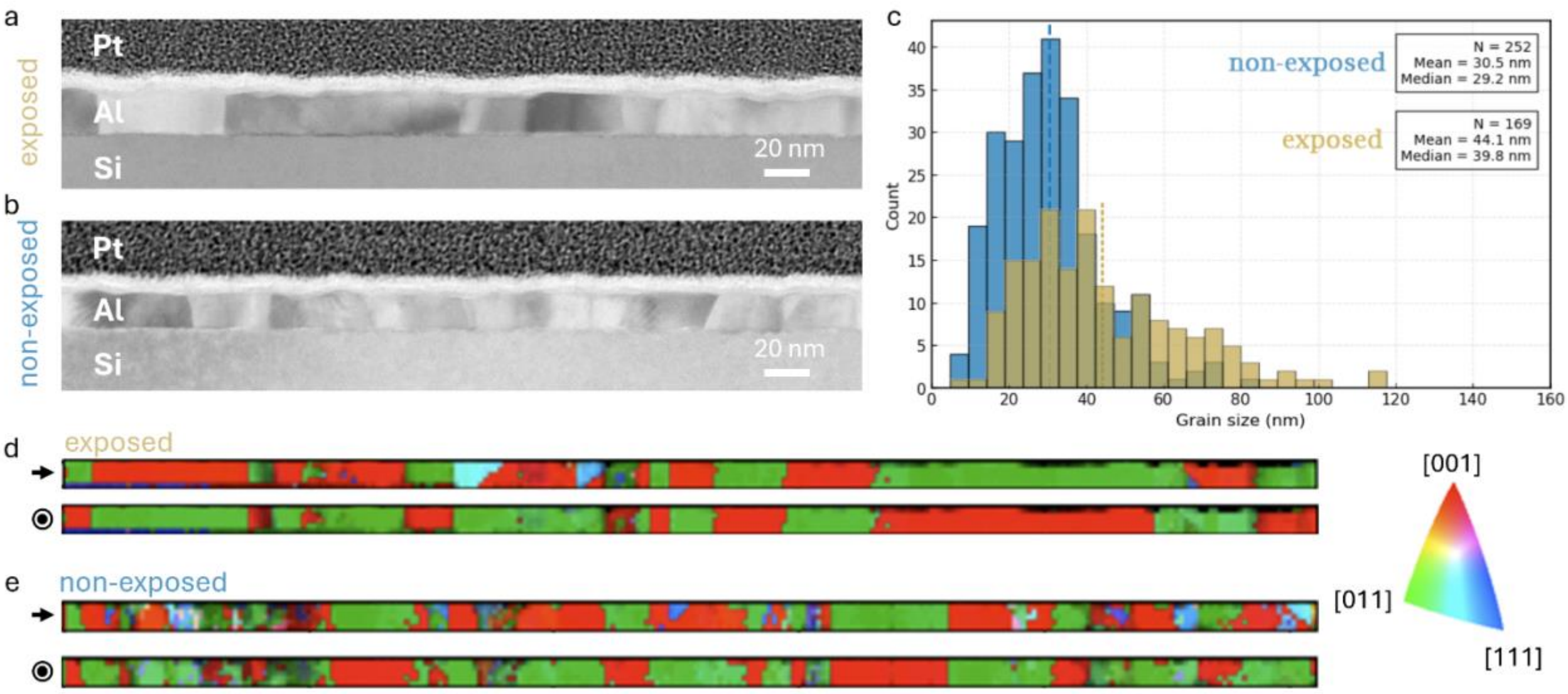

*Figure SI 4. Low magnification STEM of lamellae prepared from the exposed (a) and non-exposed (b) part of the device. Histogram showing the distribution of grain sizes in the exposed and unexposed lamellae (c). Orientation mapping on exposed (d) and non-exposed (e) lamella. The field of view is 1.2 um in both cases. The symbols on the left side of the map indicate direction. The red and green colors correspond to the [001] and [011] crystallographic families, respectively.*

### Visualisation of the grain orientation

In Figure SI 8 we schematically illustrate the orientations of Al grains in the film. Figure SI 8a corresponds to the configuration observed in the STEM image in Fig. 2b of the main text, where misfit dislocations (MDs) with a periodicity of ~3.5 nm are visible. These MDs arise from the ~5% lattice mismatch between Al and Si, which is illustrated in the "top-view" structural representation in Figure SI 8c.

In the orthogonal in-plane direction ([10-1]$_{Al}$ and [110]$_{Si}$), the lattice mismatch is significantly larger, approximately ~25% (Figure SI 8b,c), which contribute to the formation of the rectangular pattern of dislocations that is modelled below.

### Energy-dispersive X-ray spectroscopy (STEM-EDS) analysis of the Al-Si interface

The Al-Si interface after conclusion of LiPS experiments does not show any O, which would show up as a red line (Figure SI 5). The uniform O background is from oxygen exposure of the surface of the lamella during transfer from FIB to the HRTEM. The interface appears sharp, indicating that there is no significant alloying or diffusion associated with the LiPS effect.

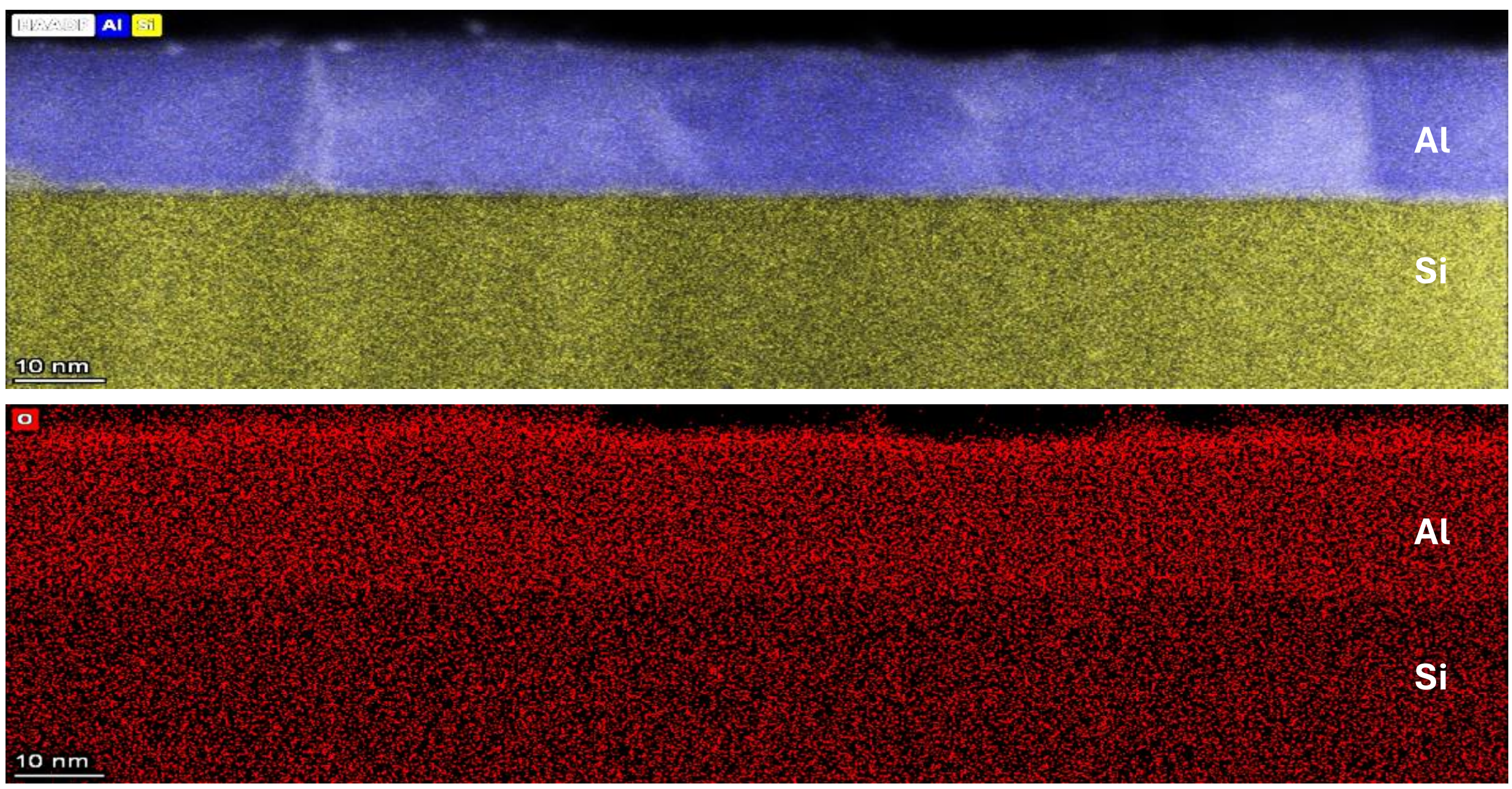

*Figure SI 5. STEM-EDS showing Al and Si signals (top) and O (bottom). The absence of a line in the bottom image indicates that there is no O layer between the Al and Si. The uniform O signal comes from residual oxide on the surface of the lamella formed during transfer from the Ar plasma milling chamber to TEM chamber in air.*

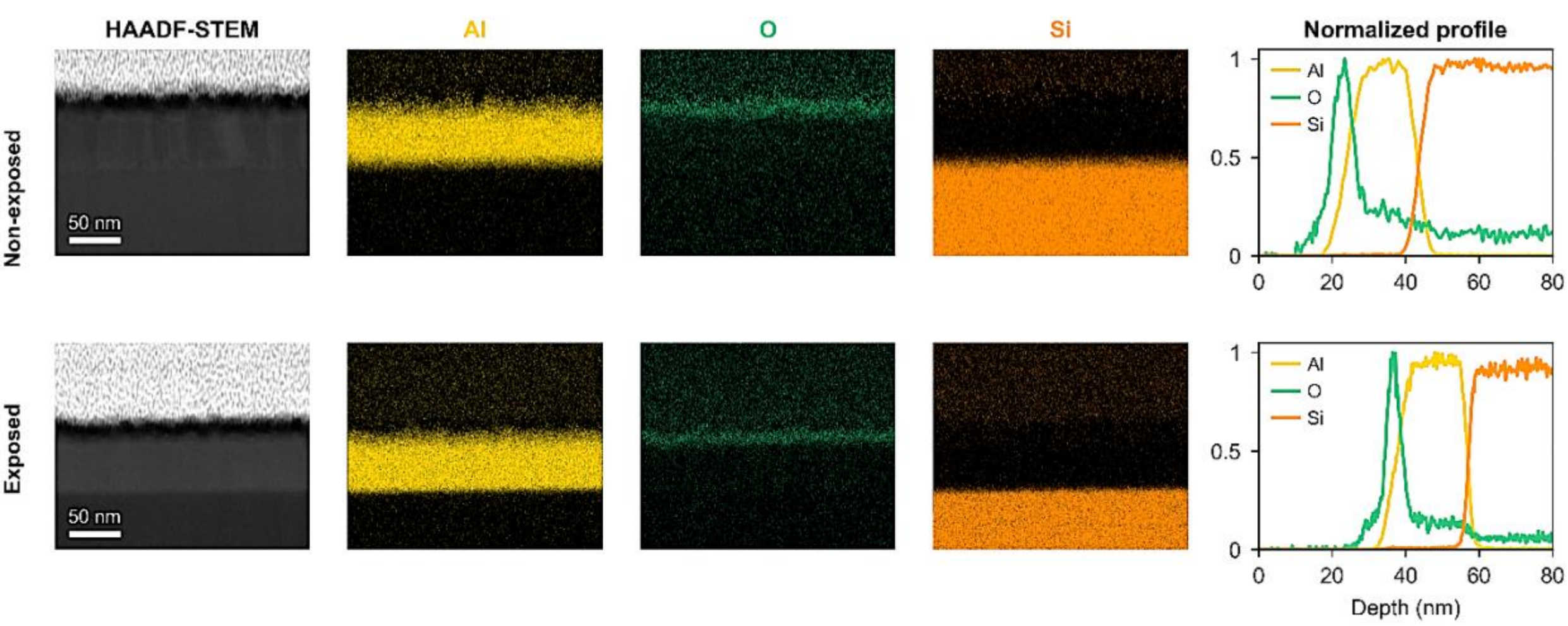

*Figure SI 6. STEM EDS for a light-exposed Al-Si cross-section (bottom) compared with a non-exposed part of the device (top). Note that the interface shows no broadening that can be attributed to alloy formation or diffusion due to the LiPS cycling at low temperatures. An oxygenated layer is observed at the top of the film, but no oxygen layer is detected at the interface which was fabricated by deposition of Al onto etched Si in an oxygen-free environment.*

## Electron energy loss spectroscopy (EELS) of the Al-Si film.

EELS spectra (Figure SI 7) are recorded on the Al-Si heterointerface, comparing light-exposed and non-exposed sections of the device after warming to room temperature. We cannot detect any alloy formation or diffusion associated with the LiPS effect: the sharpness of the interfaces is unchanged within the ~1 nm resolution of the EELS.

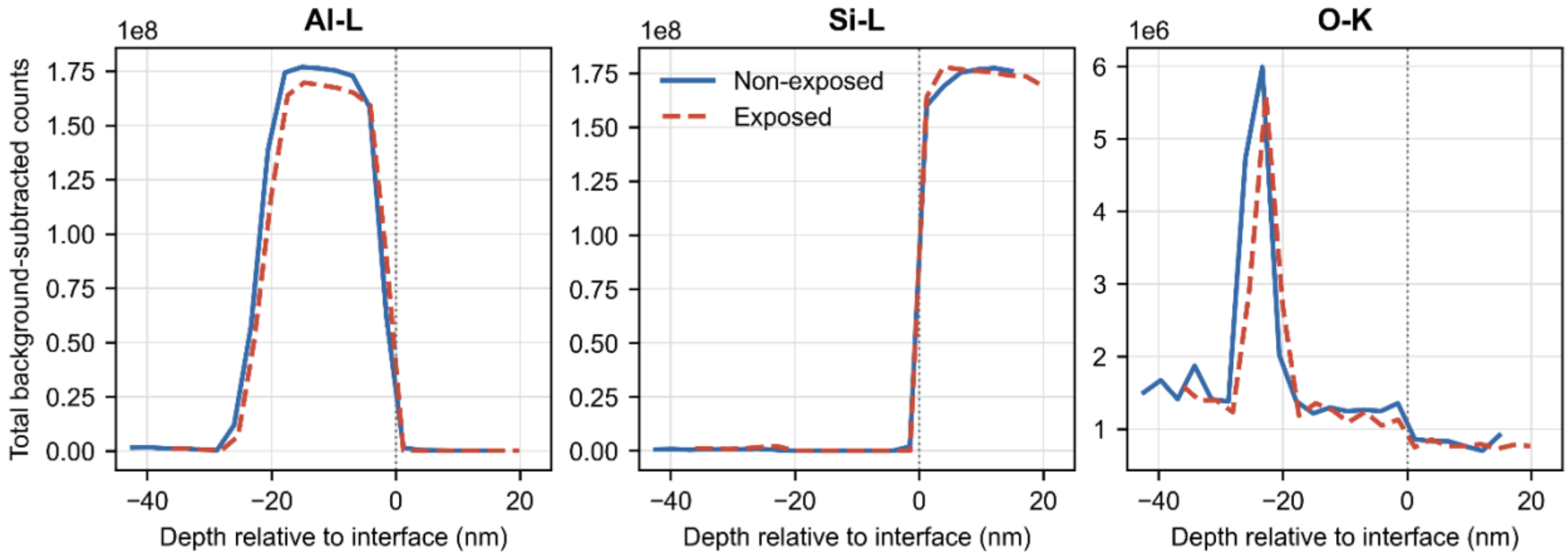

*Figure SI 7. EELS measurements for a light-exposed Al-Si cross-section (red) compared with a non-exposed part of the device (blue). The oxygen signal shows the surface of the Al is oxidized. There is an overall small O signal over the entire lamella, but there is no O at the interface (at x=0). Similarly as in EDS, the uniform O background signal comes from residual oxide on the surface formed during transfer (taking a few minutes in air) from the Ar plasma milling chamber to the STEM chamber.*

## High-magnification imaging of the Al-Si interface along the $[10\text{-}1]_{Al}$ and $[010]_{Al}$ projections.

To observe the interface along two non-equivalent orthogonal directions illustrated in Figure SI 8c, in Fig. SI 9 we compare the STEM data obtained on two grains, aligned orthogonally. In the first grain, the Al$[0\bar{1}0]$ direction is parallel to the substrate's Si$[1\bar{1}0]$ (Figure SI 8a,c). In the second grain, Al$[10\bar{1}]$ was aligned with Si$[1\bar{1}0]$ (Figure SI 8c,d). As can be observed from the corresponding FFTs (Figure SI 9b,d), in the first case, there is a small lattice mismatch of ~5%. As a result of this mismatch, misfit dislocations are observed with the periodicity of ~3.67 nm (Figure 2b-f of the main text). On the other hand, there is a large misfit of ~25% of opposite sign along Al$[10\bar{1}]$ yielding a periodicity of ~0.59 nm.

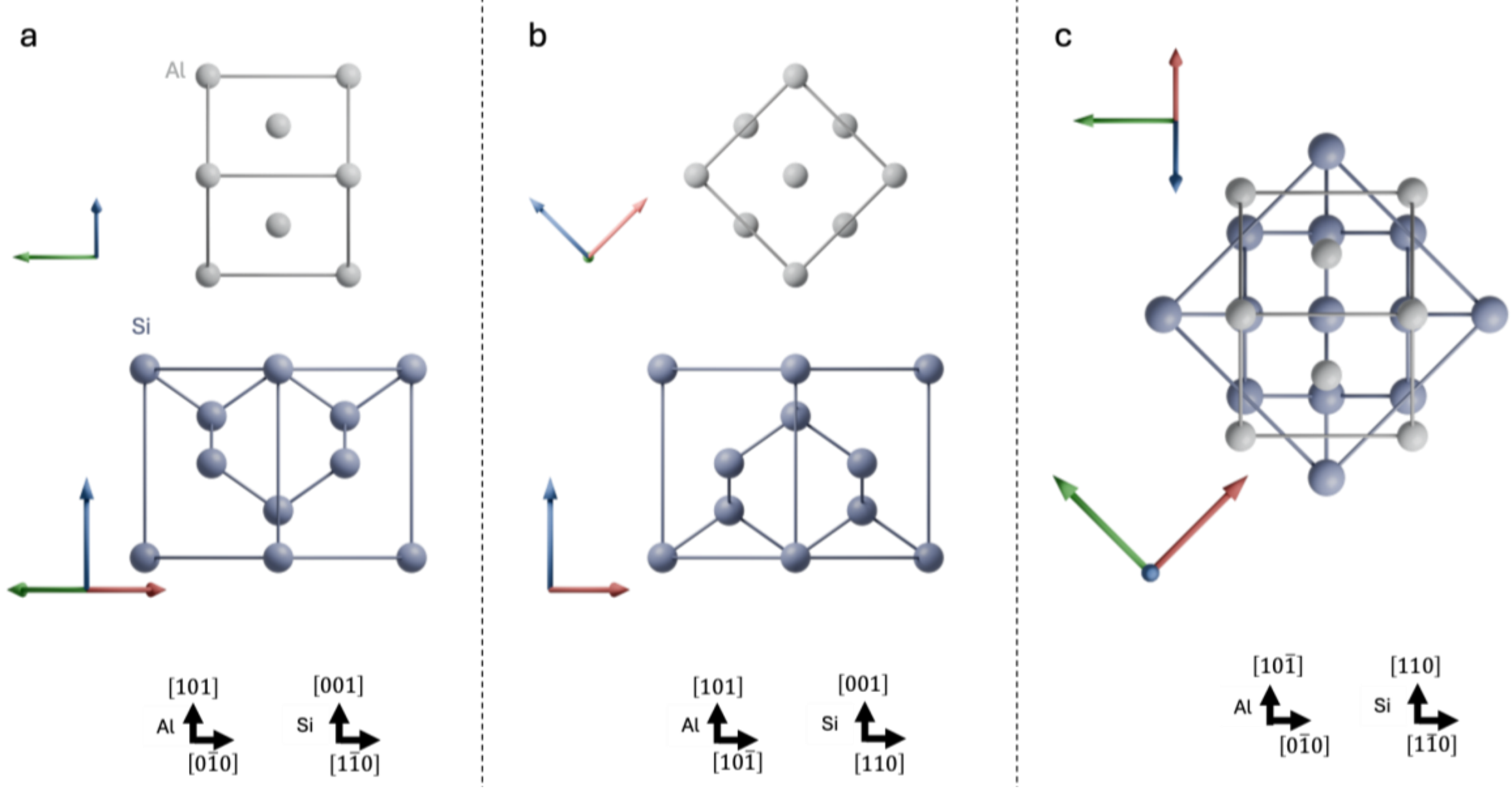

*Figure SI 8. Visualization of the Al orientation on the Si substrate as observed from three orthogonal directions. The color arrows show the main crystall directions for Al (top) and Si (bottom). Black arrows denote the crystal directions aligned along the horizontal and vertical axes of the image. The Al structure is cubic with a lattice constant is 4.05 Å, while Si is cubic with a lattice constant of 5.43 Å. The lattice mismatch at the Al [010] // Si [110] interface is 5.47%, while the lattice mismatch for the Al [101] // Si [110] interface is 25.4 %.*

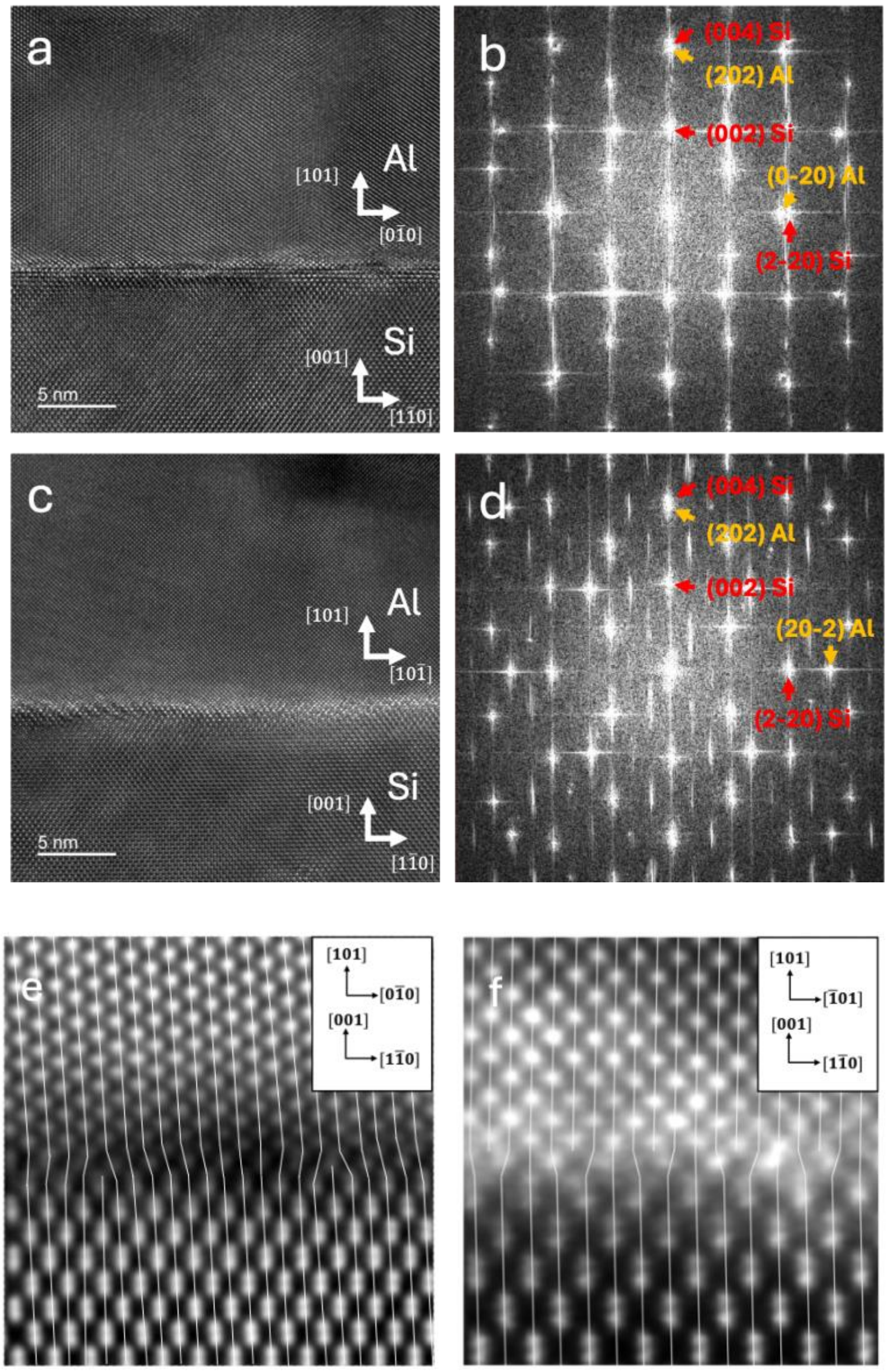

Figure SI 9. HRTEM images (a,c) and corresponding FFTs (b,d) of two differently oriented Al grains (see text). Panels e and f show the large magnification HAADF-STEM images of dislocations. The misfit dislocation period measured at room temperature after LiPS experiments is $3.67 \pm 0.49$ and $2.87 \pm 0.26$ for Al[0$\bar{1}$0] on Si[110] and $0.59 \pm 0.01$ nm for Al[10$\bar{1}$] on Si[110]. The measured dislocation period for Al[0$\bar{1}$0] on Si[110] before LiPS experiments is $3.71 \pm 0.65$ $nm$.

All the microscopy experiments are performed at room temperatures. As mentioned in the main text, and discussed below in Section 5 of the SI, resistivity measurements show that the LiPS effect is relaxed when the device is heated to room temperature. We confirm by HAADF-STEM that there is no discernible difference between the MD periodicity in either direction before and after the LiPS experiments (Figure SI 10). Majority of the $[10\text{-}1]_{Al}$ grains reveal similar MD periodicity (3.67±0.49 nm in exposed area and 3.71±0.65 nm in unexposed). However, after laser exposure we occasionally observe also grains with shorter MD periodicity (2.87 ± 0.26 nm), while without exposure such shorter periodicity was not observed. This is consistent with the model calculations of MD dynamics presented in Section 7, if we interpret the appearance of the 2.87 nm period as incomplete relaxation of the LiPS state at room temperature. It may also be an indication of a possible intermediate metastable MD Moiré state.

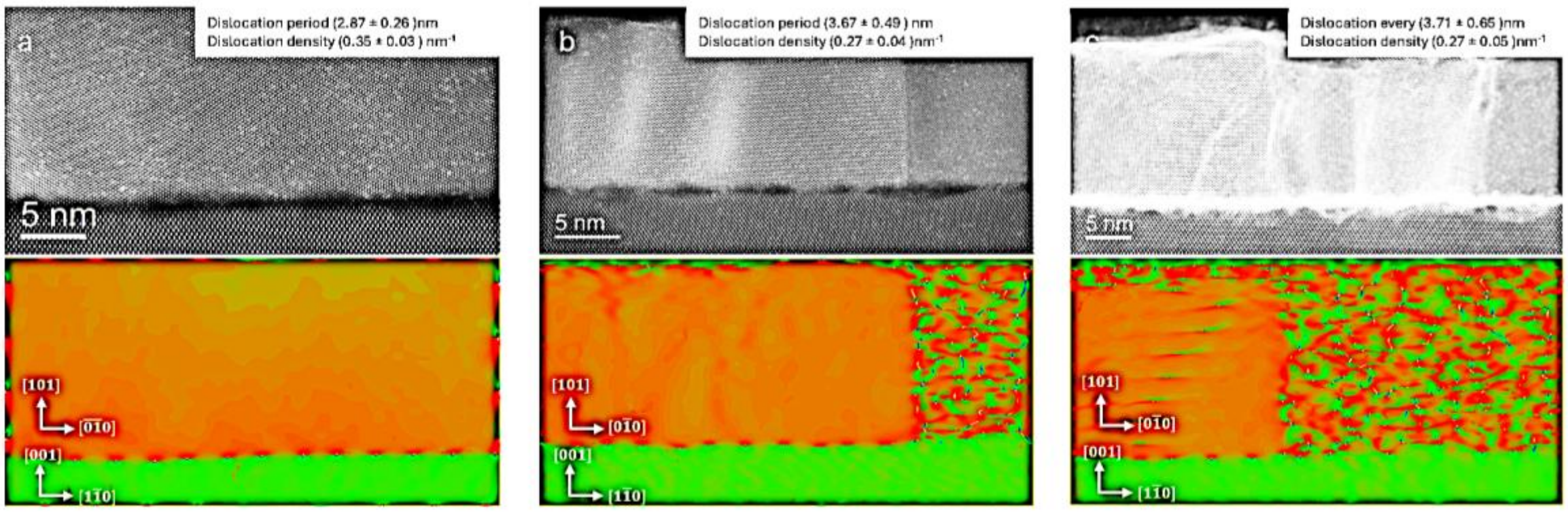

*Figure SI 10. Comparison of HAADF-STEM images and strain maps for interfaces exposed to light (a and b) and non-exposed interface (c) that underwent the same temperature cycling. The sample exposed to light shows two different MD periodicities for the same Al grain orientation ($[10\bar{1}]_{Al}$): a) occasionally 2.87 and b) more commonly 3.67 nm._Comparison of MD dislocation period of exposed (3.67 ± 0.49 nm) and non-exposed (3.71 ± 0.65 nm ) periods shown here for the [010] Al horizontal direction shows no discernible difference within the error of the measurement.*

## 4. Phenomenological model of the resistivity below $T_{c2}$ from the effective-medium approximation (EMA) model

The EMA formulates the problem of an inhomogeneous mixture of normal regions with conductivity $\sigma_n$ and superconducting puddles in the mixed phase with conductivity $\sigma_{pg}(T) \to \infty$ applicable when the superconducting state is spatially inhomogeneous. Presently, the inhomogeneity arises as a result of the inhomogeneity of the superconducting gap arising from irregular Moiré patterns and spatially irregular Brillouin zone folding (see later for a calculation of spatial inhomogeneity of the resistivity). The key variable in this simple model is the area fraction $p(T)$ occupied by the paired regions. Within this model the resistivity is controlled not only by local pairing strength (controlled by $T_c^{MF}$), but also by the connectivity of the paired regions across the sample.

For a binary mixture in 2D, the Bruggeman effective-medium equation is:

$$p(T)\frac{\big((d-1)\sigma_{pg}(T)-\sigma_e(T)\big)}{\big(\sigma_{pg}(T)+\sigma_e(T)\big)} + \big(1 - p(T)\big)\frac{\big((d-1)\sigma_n - \sigma_e(T)\big)}{\big(\sigma_n + \sigma_e(T)\big)} = 0$$

Where $d$ is the dimensionality. Using a parametrization for the superconducting fraction

$$p(T) = p_{max}\left(1 - \frac{T}{T_c^{MF}}\right)^{\beta} \quad for\, T < T_c^{MF}, \quad and \quad p(T) = 0 \quad for\, T \geq T_c^{MF}$$

the conductivity of the paired regions can be written as

$$\sigma_{pg}(T) = \sigma_n + \Delta_\sigma \left(1 - \frac{T}{T_c^{MF}}\right)^{\alpha} \quad for\, T < T_c^{MF}$$

Here $\alpha$ controls growth of local superconductivity. The resistivity is then given by $\rho_e(T) = 1/\sigma_e(T)$.

Fig. SI 11 shows the normalized resistivity $\rho/\rho_N$ as a function of temperature for two sets of parameters in 2D and in 3D. Note that the shape of the curves can easily be made to correspond to the data shown in Figs. 1h and 2a of the main text. Beyond predicting a smooth resistance curve due to inhomogeneity, the simple EMA model does not give any additional information about the system or the nature of the interactions in the system. The resistivity anisotropy for a more elaborate case of an anisotropic

Josephson-coupled Moiré MD lattice is presented in section 8, devoted to calculation of the superconducting properties of a Moiré MD lattice.

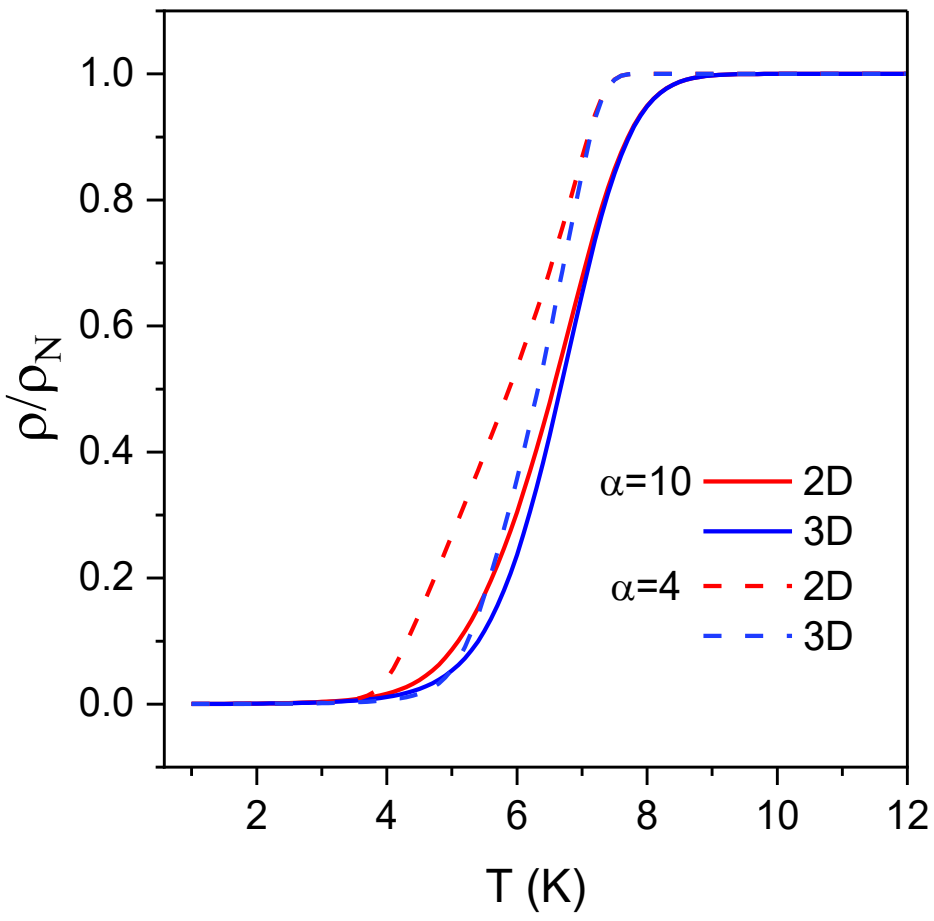

*Figure SI 11. Plots of the resistivity as a function of temperature in the EMA. 2D and 3D approximations are shown. The parameters used for the plots are: Solid curves: $T_c^{MF}$ = 12 K, $\beta = 1$; $\sigma_{PG} \to \infty$; $\alpha = 10$ ; dashed curves: $T_c^{MF}$ = 8 K, $\beta = 1$; $\sigma_{PG} \to \infty$; $\alpha = 4$.*

## 5. Switching protocol, various examples, thermal stability and reproducibility

The low-temperature superconducting state $T_c^{(1)}$ can be induced by a single high-energy pulse. A distinct resistivity downturn at low temperatures emerges at fluences ~25 mJ/cm², and becomes increasingly more pronounced with increasing pulse energy. $T_c^{(1)}$ reaches its maximum at ~50 mJ/cm². Higher pulse energies typically degrade the device, increasing normal-state resistance without further raising $T_c^{(1)}$.
Conversely, the metastable high-temperature state $T_c^{(2)}$ requires cumulative pulse exposure, which can be considered as "training". A downturn in resistance at ~8 K develops following a train of ~25-30 mJ/cm² pulses. The effect does not show any measurable dependence on the pulse repetition rates (0.2 Hz–20 kHz). In this regime, the DC heating is negligible and the sample cools back to the base temperature between the pulses. The exact shape of the R(T) curve varies between different switching attempts, suggesting inherent inhomogeneity of the system.

### Low-temperature superconductor state formation

The formation of a superconducting $T_c$ with zero resistance in the range 1.5-2.2 K is achieved by the application of a single, high-fluence laser pulse below the device damage threshold. Following pulse exposure, all tested devices exhibited the characteristic $T_c^{(1)}$ transition. However, the precise value of $T_c^{(1)}$, as determined from $\rho$(T) measurements, showed variation across devices, typically within a range of ~0.3 K. We believe this variability is primarily influenced by precise device alignment relative to the laser beam (limited by the vibrations of the cryostat) and fluctuations in the laser pulse fluence.

Furthermore, the sharpness of the $T_c^{(1)}$ transition varied across devices. Some devices exhibited a broader, less distinct transition (e.g., compare Fig. 1e in the main text with the red trace in Fig. 2a), suggesting gradual development of the high-$T_c$ state in some of the devices already after a single laser pulse.

## High-temperature superconductor formation

A "training" process involving repeated laser pulses leads to a progressively higher onset of superconductivity $T_c^{(2)}$. The number of pulses required to shift $T_c^{(2)}$ close to 8K typically exceeded several thousand. The shape of the resulting $\rho(T)$ curve in the high-$T_c$ state, as well as the magnitude of the residual resistance, showed variability between samples and was sensitive to the specific laser excitation history. Several representative $\rho(T)$ curves for different samples are presented in Figure SI 12.

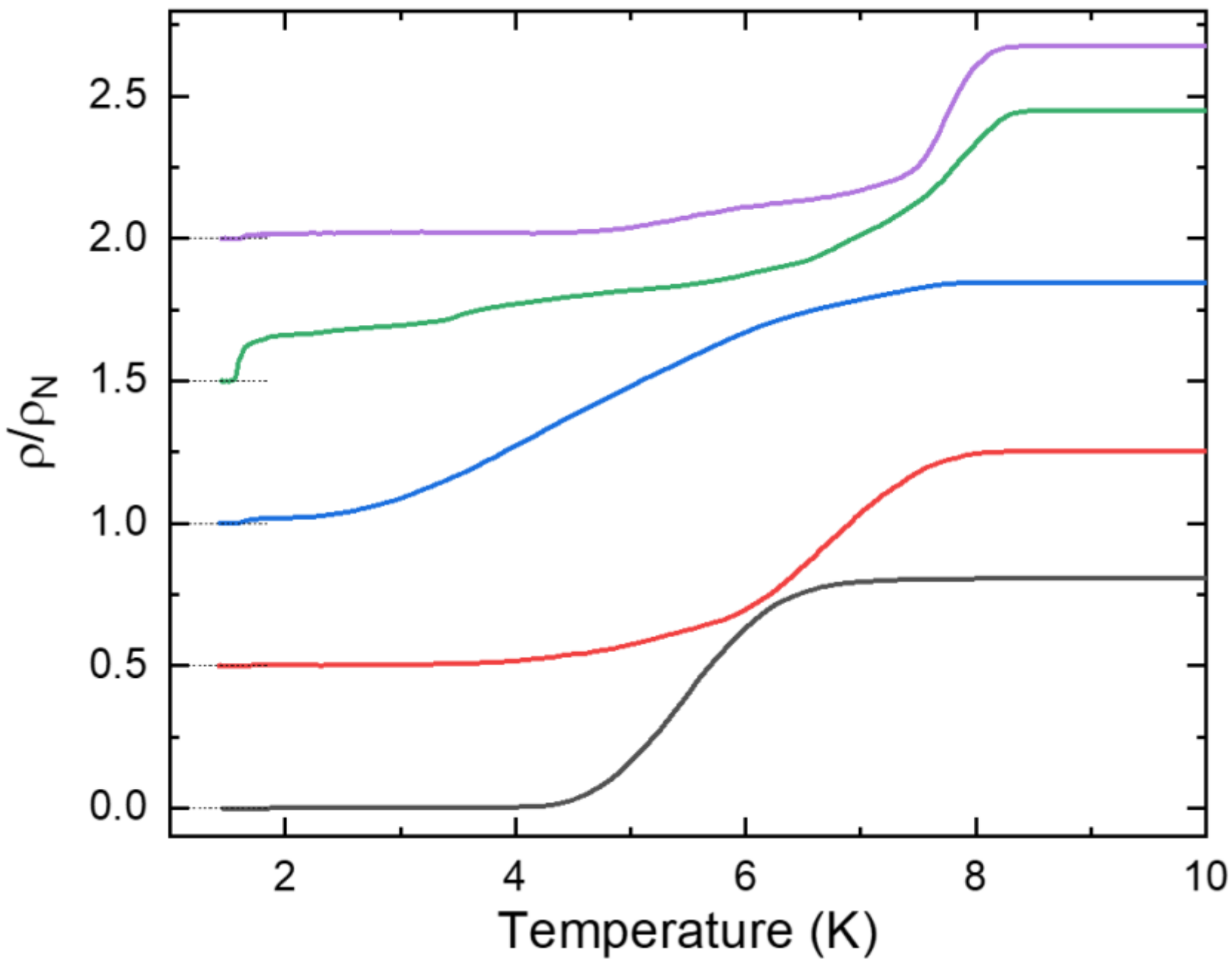

*Figure SI 12. Representative $\rho(T)$ curves obtained on 5 different devices. Note that variations are not solely caused by device-to-device variations but also by different excitation protocols. The curves are selected to showcase the diversity of observed outcomes following photoexcitation.*

## Laser erasing protocol

To erase the photoinduced state, we apply a train of more than 10 000 relatively weak laser pulses with an energy below the switching threshold (~10-20 mJ/cm$^2$) at repetition rates typically between 2 and 2000 Hz (Figure SI 13). At these repetition rates, no significant DC heating (heat accumulation) is expected, allowing the sample to cool back to the base temperature between pulses. Although erasing can also be achieved at higher repetition rates, the cooling power of the used cryostat is insufficient to compensate for the additional heat load, leading to gradual DC heating of both the sample and the sample holder.

We find that the energy window for effective erasing is relatively narrow: pulses that are too strong tend to increase $T_c^{(2)}$, whereas pulses that are too weak result in very slow relaxation. To maintain efficient erasing, a slow energy ramp envelope can be employed, in which the pulse energy is gradually reduced throughout the pulse train (Figure SI 13a). While the complete erasure of a resistance drop at low temperature was not achieved, zero resistance below $T_c^{(1)}$ is reduced to temperatures below our measurement limit of 1.4 K.

Figure SI 13b shows repetitive switching-erasing cycles indicating that the final shape of the curve is determined essentially by the erasing pulse sequence, not the initial switched state.

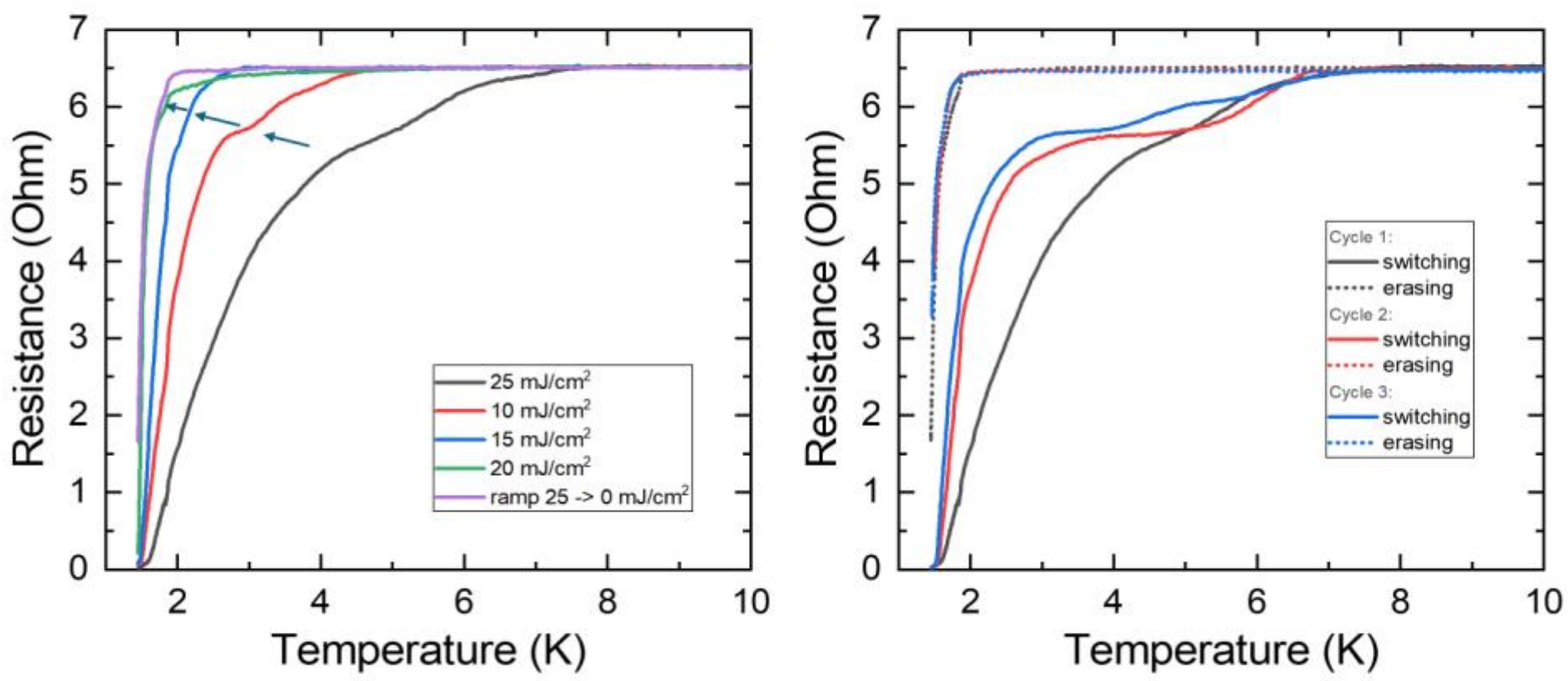

*Figure SI 13. **Erasing protocols.** The device is switched using 25 mJ/cm² pulses. Lower-fluence pulse trains (10-20 mJ/cm²) induce partial relaxation. A linear energy ramp from 25 to 0 mJ/cm² maximizes erasure efficiency. (b) Repetitive switching-erasing cycles: irrespective on the initial switched state, the shape of R(T) after erasing is determined by the erasing pulse sequence.*

## Thermal Stability of the photoinduced states

While $T_c^{(1)}$ and $T_c^{(2)}$ remain stable at low temperature on the laboratory time scale, their stability decreases as the temperature increases (Figure SI 14). Thermally cycling the sample up to 50 K, we start to observe the first signs of relaxation, which become increasingly more pronounced at higher temperatures. Elevated temperatures lead to a gradual shift of the onset temperature $T_c^{(2)}$ towards lower values until eventually the resistivity downturn shifts outside the measured temperature window. Extended time at room temperature leads to near-complete relaxation of the photoinduced state (Figure SI 14b).

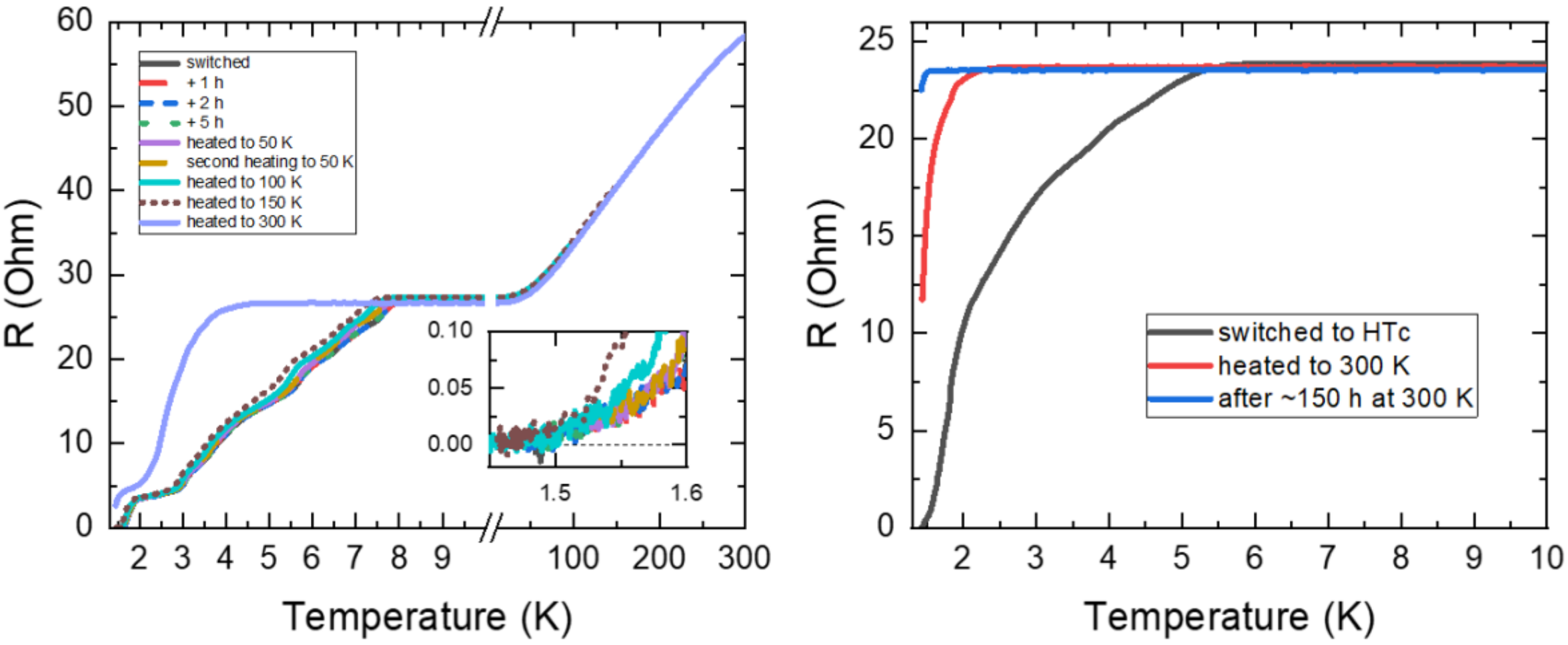

*Figure SI 14. **Thermal stability of Al-on-Si devices.** (a) No sign of relaxation of photoinduced state is observed if the sample is kept at low T<12K. A single heating cycle to 50 K induces a slight downward shift in $T_c^{(2)}$. Repeated cycling at 50 K does not lead to further change, whereas progressive heating drives continuous relaxation. (b) While short exposure to 300K, leads to partial relaxation, extended time at elevated temperatures further erase the photoinduced state.*

The relaxation rate varies between the samples and is dependent on the laser excitation history.

### Device-to-device variability

Within the present research, more than 20 devices were tested. While all devices fabricated demonstrated qualitatively consistent behavior, some quantitative differences were observed, which we attribute primarily to variations in experimental conditions, such as laser spot position and potentially subtle fabrication inconsistencies.

### $\rho(T)$ curves

All tested devices exhibited characteristic aluminum thin film resistivity, as demonstrated by consistent $\rho(T)$ behaviour (Figure SI 15). Apparent variation in resistivity between the devices are primarily attributable to minor geometric differences arising from experimental uncertainties during device fabrication (e.g., sample and mask alignment) as confirmed by SEM and TEM analysis. For resistivity calculation, nominal geometric parameters were used rather than actual for each device. However, this apparent variation does not impact the validity of the presented conclusions.

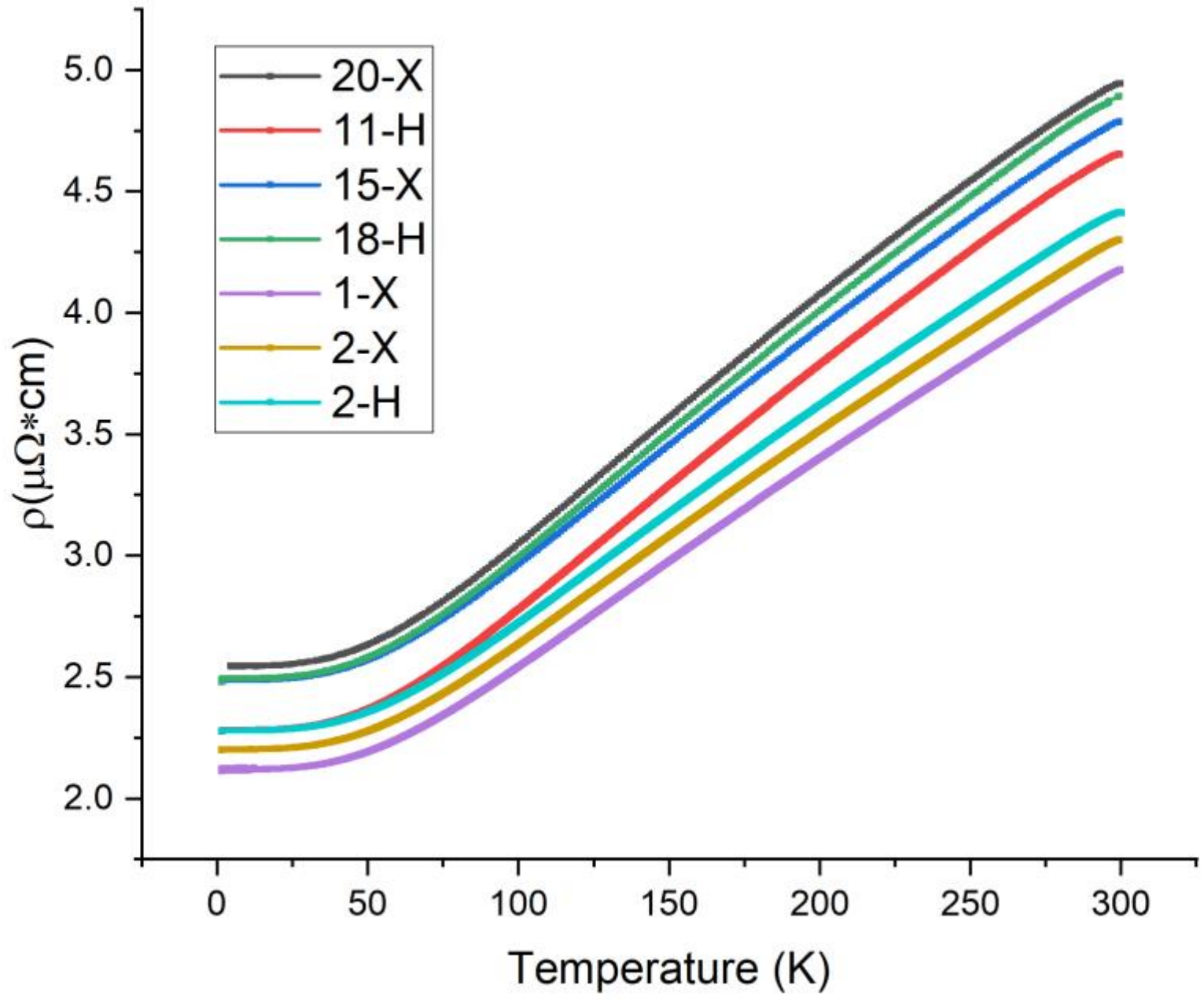

*Figure SI 15. Full $\rho(T)$ curves in the temperature range 1.4-300 K for several different devices (identified in the legend)*

## 6. Si and Al atom diffusion analysis

The mass transport after a laser pulse heats the sample is assumed to be governed by Fick's Second Law. The transient lattice temperature $T_l(t)$, leads to a diffusion coefficient $D$ which is time-dependent:

$$D(t) = D_0 \exp\left(-\frac{E_a}{k_B T_l(t)}\right)$$

The effective diffusion product $(Dt)_{eff}$ after a time $t_{end}$ is obtained by integration:

$$(Dt)_{eff} = \int_0^{t_{end}} D\,(T_l(\tau))\,d\tau$$

The resulting normalized concentration profile $C(x)$ as a function of depth $x$ is:

$$C(x) = C_0 \exp\left(-\frac{x^2}{4(Dt)_{eff}}\right)$$

The following table summarizes the diffusion parameters used for the Al-Si system diffusion calculation shown in Figure Si 16.

*Arrhenius parameters for the Al-Si system.*

| **Parameter** | **Si in Al** | **Al in Si** |
|---|---|---|
| Pre-exponential $D_0$ ($m^2/s$) | $1.38 \times 10^{-5}$ | $8.0 \times 10^{-4}$ |
| Activation Energy $E_a$ (eV) | 1.24 | 3.47 |

In the simulation, the lattice temperature peaks at $T_l \sim 900$ K within first 10 ps. This value is set by the laser fluence. Afterwards, the heat is transferred into the bulk and the temperature of the Al film and the Al-Si interface quickly drops. Due to the high diffusion activation energy for Al in Si, the diffusion of Aluminum into Silicon is negligible on this timescale. Silicon diffusion into Aluminum is faster but still remains confined to the sub-nanometer scale ($< 0.1$ nm) on the timescale of the thermal pulse (Figure SI 16). This implies that Al and Si interdiffusion effects are fundamentally limited at the interface.

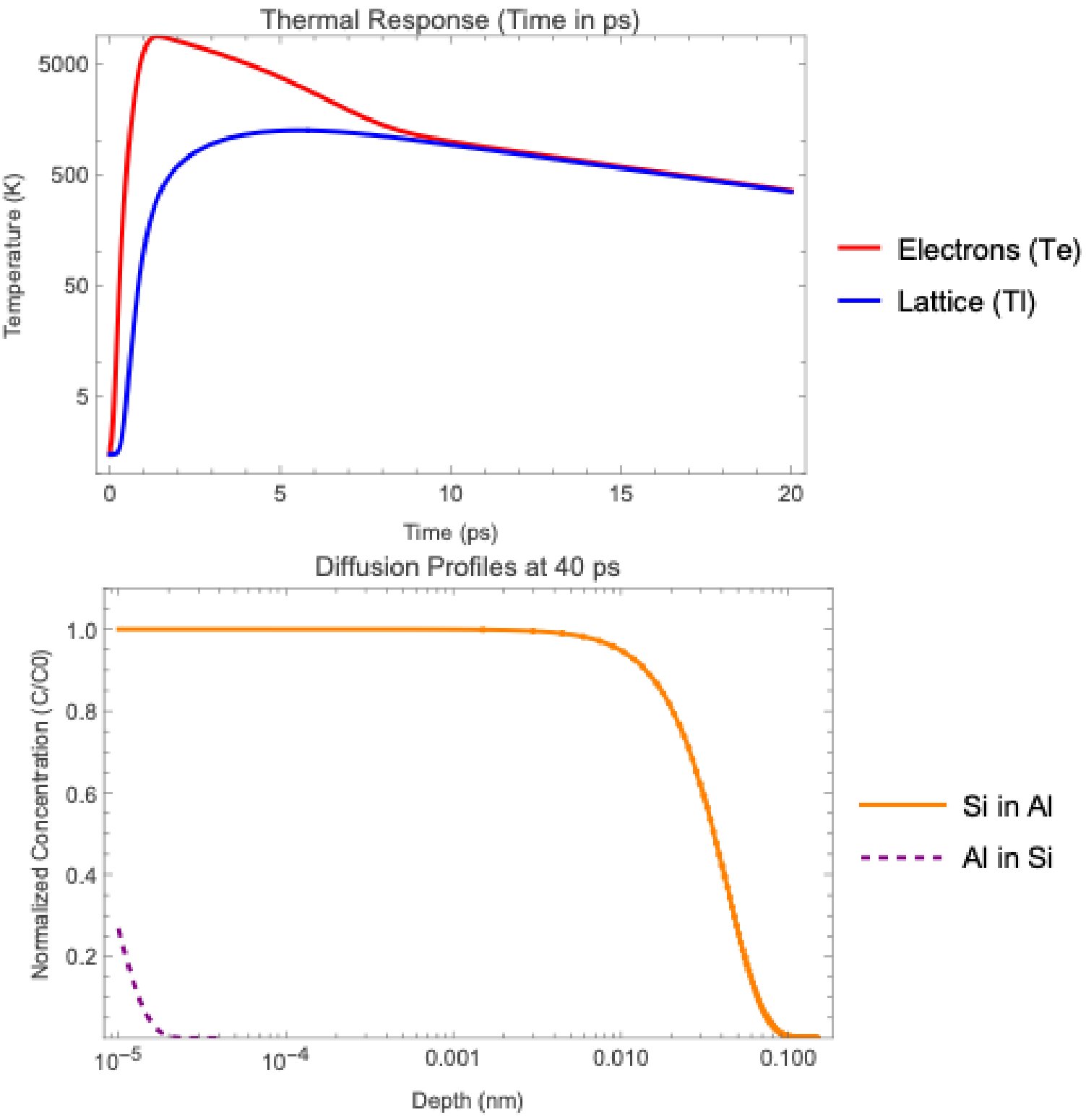

*Figure SI 16. The thermal response (top) and corresponding diffusion profiles for Si in Al and Al in Si (bottom). Note that the majority of diffusion takes place in the first 10 ps.*

## 7. Dynamical Model for Misfit Dislocations in Al/Si Interfaces

### Discrete Frenkel-Kontorova (FK) Model

Discommensuration between the Al and Si lattice can be modelled using the Frenkel-Kontorova model[7]. The interface is initially modeled as a discrete chain of $N$ Aluminum (Al) atoms coupled by harmonic springs and subjected to the periodic potential of the Silicon (Si) substrate. The Hamiltonian $H$ for the system is given by:

$$H = \sum_i \left[ \frac{1}{2} m\dot{u}_i^2 + \frac{1}{2} k(u_{i+1} - u_i - a_{Al})^2 + V_0 \left( 1 - \cos\left( \frac{2\pi u_i}{a_{Si}} \right) \right) \right]$$

Where $u_i$ is the position of the $i$-th Al atom, $k$ represents the Al-Al bond strength (spring constant), $a_{Al}$ and $a_{Si}$ are the lattice constants of the film and substrate, respectively and $V_0$ is the depth of the periodic substrate potential. In this model, misfit dislocations are treated as discommensurations—topological deffects where the phase of the Al lattice shifts relative to the Si lattice.

### Dynamical Evolution under Thermal Transients

To account for the response to a transient temperature pulse $T(t)$, we employ Langevin dynamics to simulate the atomic positions:

$$m\frac{d^2u_i}{dt^2} = -\gamma\frac{du_i}{dt} - \frac{\partial V_{total}}{\partial u_i} + \xi_i(t)$$

Here the damping coefficient $\gamma$ represents phonon drag and energy dissipation. The temperature pulse $T(t)$ facilitates the crossing of the Peierls-Nabarro barrier, allowing dislocations to migrate. The stochastic force $\xi_i(t)$ represents thermal noise, satisfying the fluctuation-dissipation relation where the variance is proportional to the instantaneous temperature $T(t)$.

For spatiotemporal analysis, the discrete model can be mapped to the continuous Sine-Gordon field equation. The displacement field $\phi(x,t)$ represents the local phase shift:

$$\frac{\partial^2\phi}{\partial t^2} - c_0^2\frac{\partial^2\phi}{\partial x^2} + \omega_p^2\sin(\phi) = -\Gamma\frac{\partial\phi}{\partial t} + \eta(x,t)$$

In this continuum limit the spatial derivative $\frac{\partial\phi}{\partial x}$ corresponds to the local misfit strain. Discommensurations appear as solitonic topological defects (high-strain bands). The pulse triggers a Commensurate-Incommensurate (C-IC) transition, where the system relaxes toward a new equilibrium density of dislocations.

Post-pulse dynamics are characterized by long-term relaxation: Elastic relaxation is caused by residual strain which drives the movement of dislocations after the pulse ends. An important effect is soliton repulsion, where discommensurations exert repulsive forces on one another, leading to a more uniform spatial distribution. Finally, thermal excitation allows for sub-threshold hopping over pinning sites (thermal creep). These effects are not included in the model.

For 1D model calculation, the lattice misfit along Al$[0\bar{1}0]$ / Si$[1\bar{1}0]$ was used. The atoms are considered to be elastically coupled to the two nearest neighbours and are moving in a sinusoidal potential of the Si lattice. The results of modelling are shown in Fig. 2g of the main text.

For 2D case, we use the square potential of Si(001) substrate. The misfit along Al$[0\bar{1}0]$ and Al$[10\bar{1}]$ are calculated separately. The Al atoms are considered to be elastically coupled to the nearest neighbours along two orthogonal crystallographic directions with two different coupling constants. A shear force to the next-nearest neighbours ("diagonal" atoms) is added to the Hamiltonian. The results are shown in Fig 2h of the main text and in Figure SI 17.

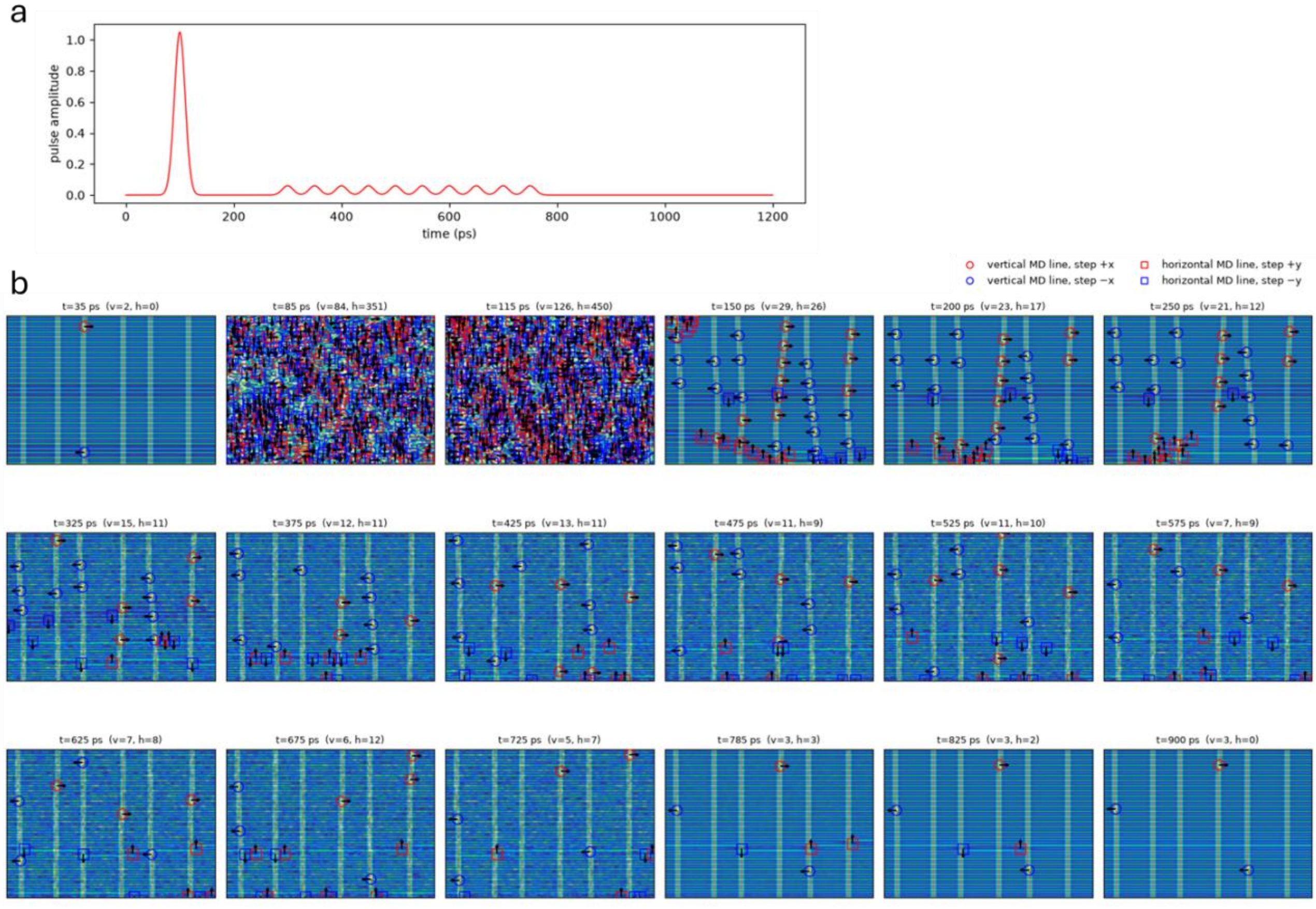

*Figure SI 17. Pulse sequence applied to the sample (a) and (b) the resulting pattern of kinks at selected times along the timeline shown in a).*

The difference between the 1D and 2D models is the robustness of the MD lines: while in 1D case the density of MDs can be easily manipulated by the pulse, in 2D pattern is more strongly pinned due to the presence of topological defects – *kinks* at the crossings.

Importantly, in 2D model we observe an increased number of kinks on the MD lines after the strong pulse. Subsequent weaker pulses cause gradual pairwise annihilation of the kinks of opposite topological charge, consistent with the E (erase) process. The kink density in response to W and E sequences is shown in the timeline plot Fig. SI 18.

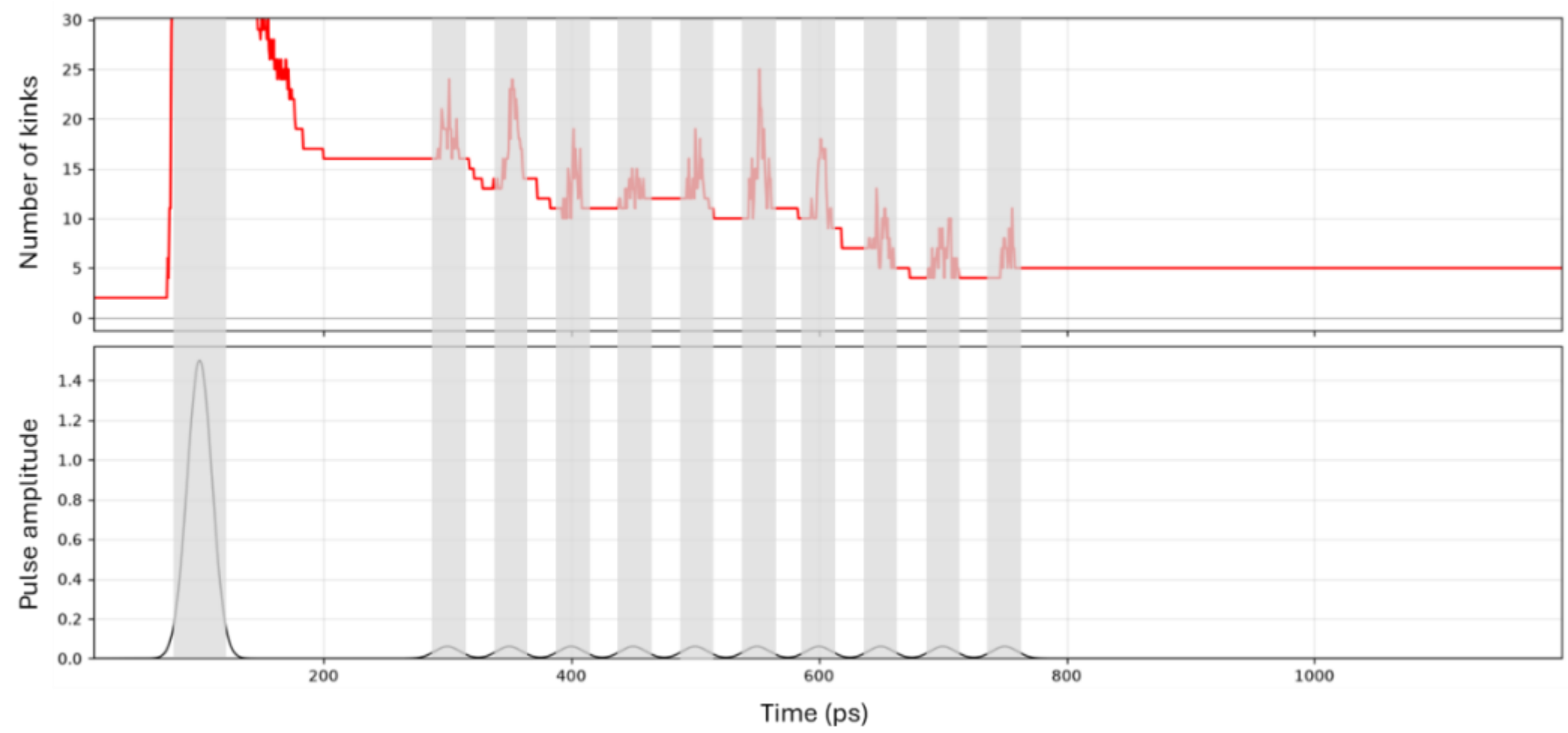

*Figure SI 18. Time evolution of the number of kinks, appearing in the MD lines running along Al*$[10\bar{1}]$ *direction. After strong laser pulse the number of kinks significantly increases. A series of weak pulses cause gradual annealing of kinks. The shaded areas mask time during the pulses.*

## 8. Calculation of the spatial variation of the superconducting gap for a discommensuration strain-warped Al(101)/Si(001) Moiré Interface.

The main idea is that misfit dislocations which form a rectangular Moiré pattern in the plane of the interface cause zone folding and the formation of minibands. The strain map shown in Fig. 2d of the main text shows that the strains originating from MDs propagate more than 5 nm both into the Al and Si. For Al (101) (fcc, $a_{\mathrm{Al}} \approx 4.05$ Å) and Si (001) (diamond cubic, $a_{\mathrm{Si}} \approx 5.43$ Å), a rectangular Moiré superlattice naturally emerges due to the $\approx 5\%$ (along Al$[0\bar{1}0]$//Si$[1\bar{1}0]$) and $\approx 25\%$ (along Al$[10\bar{1}]$//Si$[011]$).

While bulk Aluminum behaves as a conventional, weak-coupling BCS superconductor, the interfacial Moiré envelope folds the original atomic bands into a heavily compressed Moiré Brillouin zone (MBZ). This results in the formation of ultra-flat minibands where kinetic energy is severely quenched and the local density of states (DOS) is strongly enhanced. In the calculation below we calculate the spatial variation of the superconducting gap due to the Moiré MD lattice and the resulting anisotropy of the resistivity for the present case based on the effective tight-binding Hamiltonian (2) given in the main text.

### Phase-Warped Disordered Moire Superlattice

We consider a two-dimensional tight-binding square lattice $(N_x \times N_y)$ subject to an anisotropic, spatially disordered Moire potential $V_{\mathrm{Moire}}(\mathbf{r})$. To model structural lattice distortions and period fluctuations, the physical coordinates $(x, y)$ are transformed via a spatially correlated Gaussian-filtered random displacement field $\boldsymbol{\eta}(x, y) = (\eta_x, \eta_y)$:

$$\begin{aligned} \tilde{x}_i &= x_i + \delta_x\, \eta_x(x_i, y_j), \\ \tilde{y}_j &= y_j + \delta_y\, \eta_y(x_i, y_j), \end{aligned}$$

where $\delta_x$ and $\delta_y$ quantify the spatial disorder amplitude along the inter-river ($x$) and intra-river ($y$) directions, respectively. The effective Moire potential is evaluated at the warped coordinates:

$$V_{\mathrm{Moire}}(x_i, y_j) = V_0\left[\cos(k_x \tilde{x}_i) + \alpha \cos(k_y \tilde{y}_j)\right],$$

where $k_x = 2\pi/D_x$ and $k_y = 2\pi/D_y$ define the anisotropic Moire periodicity ($D_x \gg D_y$).

**BdG Hamiltonian and Self-Consistent Order Parameter**

The tight-binding Hamiltonian in the Nambu basis $\Psi_i = (c_{i\uparrow}, c_{i\downarrow}^\dagger)^T$ is given by:

$$\mathcal{H}_{\text{BdG}} = \begin{pmatrix} \mathbf{H}_0 & \mathbf{\Delta} \\ \mathbf{\Delta}^* & -\mathbf{H}_0^* \end{pmatrix},$$

where the single-particle matrix elements $(\mathbf{H}_0)_{ij}$ account for anisotropic hopping and the local disordered potential:

$$(\mathbf{H}_0)_{ij} = [V_{\text{Moire}}(\mathbf{r}_i) - \mu]\delta_{ij} - t_0 \sum_{\langle i,j \rangle} \delta_{\langle i,j \rangle} - t_x{}' \sum_{\langle\langle i,j \rangle\rangle_x} \delta_{\langle\langle i,j \rangle\rangle} - t_y{}' \sum_{\langle\langle i,j \rangle\rangle_y} \delta_{\langle\langle i,j \rangle\rangle}.$$

Here, $t_0$ is the nearest-neighbor hopping integral, while $t_x{}'$ and $t_y{}'$ denote the anisotropic next-nearest-neighbor (NNN) hopping parameters that govern diagonal kinetic coupling across and along the channels.

Diagonalization of $\mathcal{H}_{\text{BdG}}$ via the Bogoliubov transformation yields the positive eigenvalues $E_n > 0$ and eigenvectors $(u_n(\mathbf{r}_i), v_n(\mathbf{r}_i))^T$:

$$\sum_j \begin{pmatrix} (\mathbf{H}_0)_{ij} & \Delta_i \delta_{ij} \\ \Delta_i^* \delta_{ij} & -(\mathbf{H}_0^*)_{ij} \end{pmatrix} \begin{pmatrix} u_n(\mathbf{r}_j) \\ v_n(\mathbf{r}_j) \end{pmatrix} = E_n \begin{pmatrix} u_n(\mathbf{r}_i) \\ v_n(\mathbf{r}_i) \end{pmatrix}$$

The local superconducting order parameter $\Delta(\mathbf{r}_i)$ is computed self-consistently using Picard iteration until convergence ($\parallel \mathbf{\Delta}^{(k+1)} - \mathbf{\Delta}^{(k)} \parallel < 10^{-4}$):

$$\Delta(\mathbf{r}_i) = U_{\text{pair}} \sum_{E_n > 0} u_n(\mathbf{r}_i) v_n^*(\mathbf{r}_i) \tanh\left(\frac{E_n}{2k_B T}\right),$$

where $U_{\text{pair}}$ represents the effective attractive pairing interaction strength. Fig. SI 19 shows 2D maps of the Moiré potential $V_{Moiré}(\boldsymbol{r})$ and the resulting superconducting gap $|\Delta(\boldsymbol{r})|$ for moderate disorder $\delta_x = \delta_y = 0.15$. For illustration purposes two different MD lattice periods are presented: a) 2.0 x 3.78 nm, where the rectangular lattice is visible, and b) 0.59 x 3.78 nm where the anisotropy leads to strongly one-dimensional 'rivers' of pairing potential. The latter corresponds to the MD lattice in the present case.

a

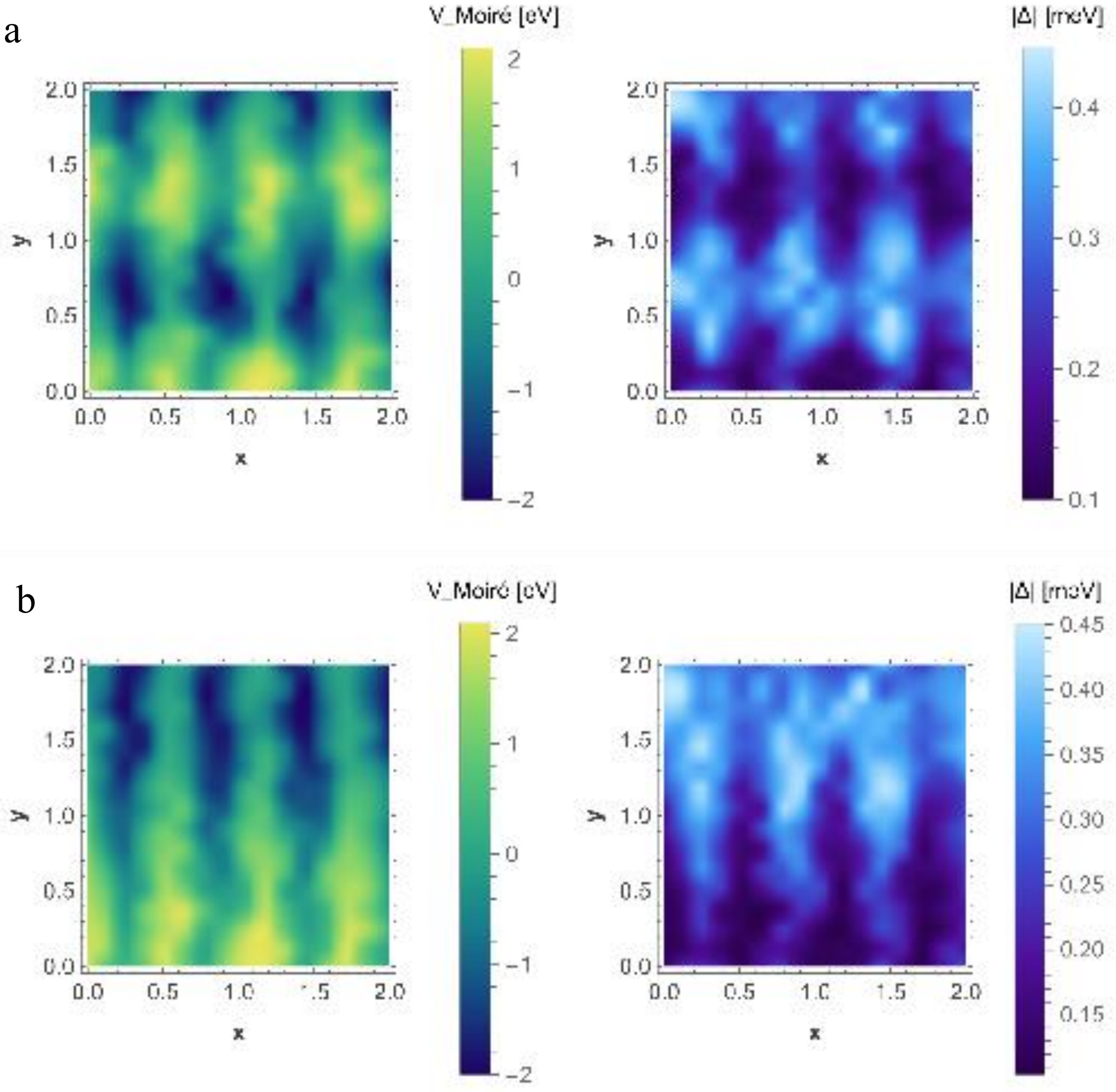

*Figure SI 19. The paring potential and the superconducting gap calculated for an anisotropic disordered Moiré lattice of misfit dislocations. a) small anisotropy: MD lattice periodicity 2.0 x 3.78 nm, and b) large anisotropy, relevant to the Al-Si interface: lattice periodicity 0.59 x 3.78 nm.*

The calculated gap function $\Delta(x)$ shows that Cooper pairs reside preferentially inside the deeply trapped potential valleys where the local density of states is maximized. Conversely, in regions where the Moiré periodicity compresses, the elevated NNN hopping dispersion suppresses the local pairing magnitude, creating localized weak-link constrictions.

## 9. Macroscopic Josephson Junction Array (JJA) Transport Model

The microscopic spatial solution $|\Delta(\mathbf{r}_i)|$ derived from the BdG solver above is mapped onto a macroscopic 2D network of Josephson junctions. The Josephson energy coupling $J_{ij}(T)$ between adjacent lattice nodes $i$ and $j$ follows the Ambegaokar-Baratoff relation[8] adjusted for local gap variation:

$$J_{ij}(T) = \frac{\pi\hbar}{4e^2 R_{N,ij}} \bar{\Delta}_{ij}(T) \tanh\left(\frac{\bar{\Delta}_{ij}(T)}{2k_B T}\right),$$

where $\bar{\Delta}_{ij}(T) = \frac{1}{2}\left(|\Delta(\mathbf{r}_i)| + |\Delta(\mathbf{r}_j)|\right)$ is the mean order parameter across the link, and $R_{N,ij} \in \{R_{N,x}, R_{N,y}\}$ represents the directional normal-state link resistance ($R_{N,x} \gg R_{N,y}$).

At finite temperatures $T < T_c$, phase fluctuations generate finite dissipation across the links. According to the Ambegaokar–Halperin theory for thermally activated phase slips[9], the effective link resistance $R_{ij}(T, \mathbf{r})$ is given by:

$$R_{ij}(T) = R_{N,ij} \left[ I_0\left(\frac{J_{ij}(T)}{2k_B T}\right) \right]^{-2},$$

where $I_0(z)$ is the modified Bessel function of the first kind of order zero.

To evaluate the directional macroscopic resistivities $\rho_x(T)$ and $\rho_y(T)$, we formulate the discrete Kirchhoff node voltage equations[10] over the network matrix $\mathbf{G}$:

$$\sum_{j \in \text{neighbors}(i)} G_{ij}\,(V_i - V_j) = I_i^{\text{ext}},$$

where the inter-node conductances are $G_{ij} = 1/R_{ij}(T)$.

Transverse transport ($\rho_x$): A total current $I_{\text{ext}}$ is injected at the boundary $x = 0$ ($i = 1$) and extracted at $x = L_x$ ($i = N_x$), subject to Dirichlet reference boundary conditions at a reference node.

Longitudinal transport ($\rho_y$): The current $I_{\text{ext}}$ is injected at $y = 0$ ($j = 1$) and extracted at $y = L_y$ ($j = N_y$).

Solving the linear system $\mathbf{GV} = \mathbf{I}$ yields the spatial voltage distribution $\mathbf{V}$. The effective resistivities are then evaluated as:

$$\rho_x(T) = \frac{\langle V(1,j)\rangle_j - \langle V(N_x,j)\rangle_j}{I_{\text{ext}}}, \qquad \rho_y(T) = \frac{\langle V(i,1)\rangle_i - \langle V(i,N_y)\rangle_i}{I_{\text{ext}}}.$$

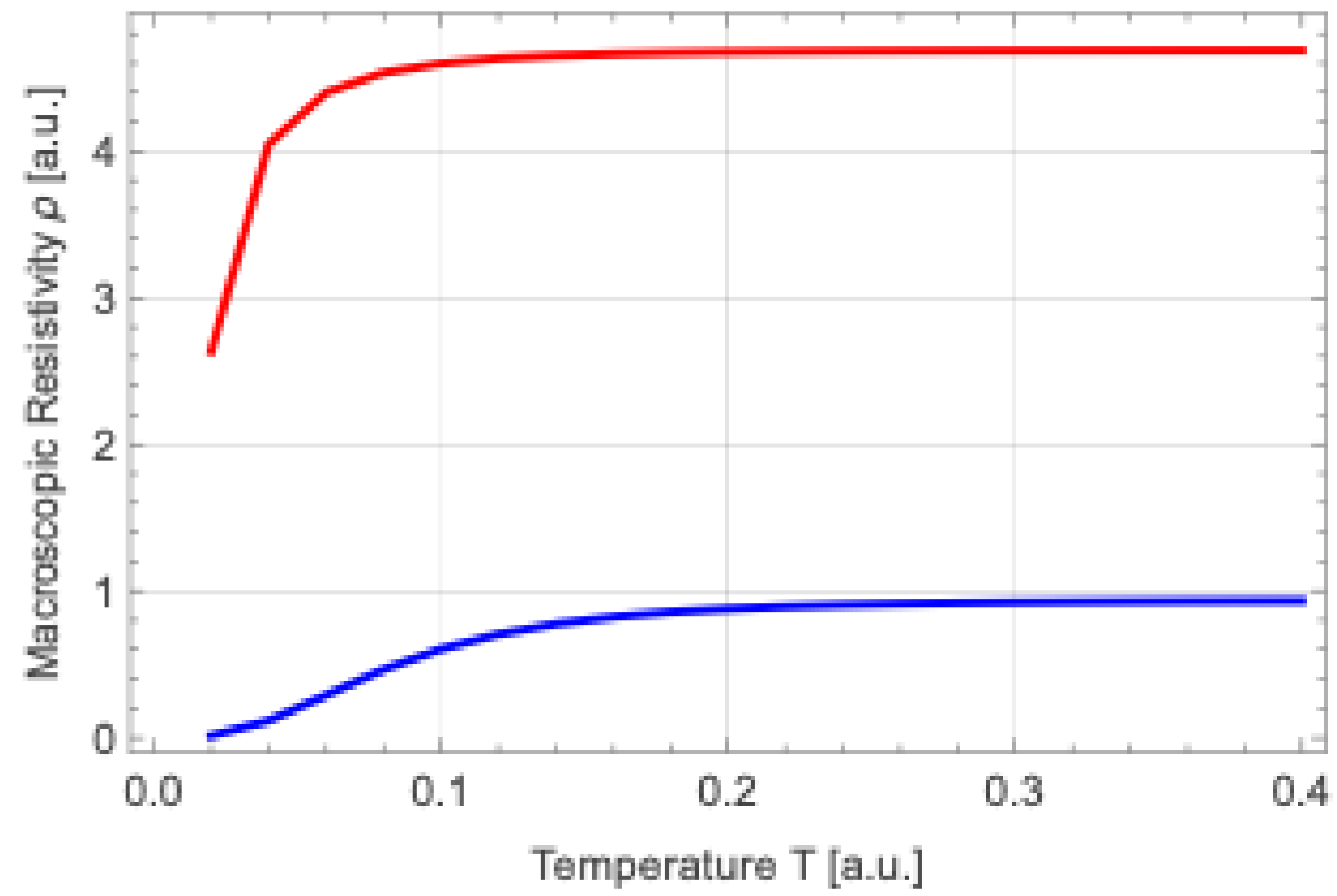

*Figure SI 20 Macroscopic resistivity calculated for an anisotropic Moiré lattice: along the 'rivers' $\rho_y$ (blue), across the 'rivers': $\rho_x$ (red).*

The model for thermally activated phase slips on a rectangular Moiré lattice gives qualitatively similar resistivity curves as the EMA calculation (section 4), even though the EMA calculation of resistivity assumes isotropic superconducting puddles, rather than quasi one dimensional 'rivers' in the anisotropic Moiré model. The validity of the presented anisotropic phase-slip model calculation can be tested by measurement of the anisotropy of the resistivity on a single Al domain on Si by STM after exposure to a laser pulse.